\pdfoutput=1
\documentclass[cernpreprint,texlive=2016,txfonts,subfigure=true,texmf,UKenglish]{atlasdoc} 
\usepackage{atlaspackage}
\usepackage{atlasbiblatex}
\usepackage{atlasphysics}

\usepackage{paper-defs}

\AtlasTitle{Searches for the $Z\gamma$ decay mode of the Higgs boson and for new high-mass resonances in $pp$ collisions at $\sqrt{s}$ = 13~\TeV\ with the ATLAS detector}

\author{The ATLAS Collaboration}

\PreprintIdNumber{CERN-EP-2017-095}

\AtlasJournal{JHEP}
\AtlasJournalRef{JHEP 10 (2017) 112}
\AtlasDOI{10.1007/JHEP10(2017)112}

\AtlasAbstract{%

This article presents searches for the \Zg decay of the Higgs boson and for narrow
high-mass resonances 
decaying to $Z\gamma$, exploiting $Z$ boson decays to pairs of electrons or muons.
The data analysis uses \lumithirteentev~\invfb~of $pp$ collisions at \sqtt~recorded
by the ATLAS detector at the CERN Large Hadron Collider. The data are found to be
consistent with the expected Standard Model background. 
The observed (expected -- assuming Standard Model $pp\to H\to Z\gamma$ production and decay) 
upper limit on the production cross section times the branching ratio for $pp\to H\to Z\gamma$ is 
 \mhObserved~(\mhExpectedSMHiggs) times the Standard Model prediction at the 95\% 
confidence level for a
Higgs boson mass of 125.09~\GeV. In addition, upper limits are set on the production
cross section times the branching ratio as a function of the mass of a narrow resonance between 
 250~\GeV\ and 2.4~\TeV, assuming spin-0 resonances produced via gluon--gluon fusion, 
and spin-2 resonances produced via gluon--gluon or quark--antiquark initial states. For high-mass spin-0 resonances, 
the observed (expected) limits vary between \highmasslimitlowmX~(\highmasslimitlowmXexp) 
and \highmasslimithighmX~(\highmasslimithighmXexp) for 
the mass range from 250~\GeV\ to 2.4~\TeV\
at the 95\% confidence level.
}

\hypersetup{pdftitle={ATLAS document},pdfauthor={The ATLAS Collaboration}}

\begin{document}

\maketitle

\tableofcontents

\section{Introduction}
\label{sec:intro}
Since the observation of a Higgs boson ($H$) by the ATLAS and CMS 
collaborations~\cite{atlas-higgs, cms-higgs}, its properties have been measured and
presented in subsequent publications. Its mass was determined
to be $\mH = 125.09 \pm 0.21\mathrm{(stat)} \pm 0.11\mathrm{(syst)}$~\GeV~\cite{higgs-mass}.
The couplings to Standard Model (SM) elementary particles were measured and confirmed to be consistent with 
the predictions for a SM Higgs boson within the present 
uncertainties~\cite{atlas-couplings,cms-couplings,atlas-cms-combined}, 
and alternative spin and $CP$ hypotheses were rejected in favour of the SM 
hypothesis~\cite{atlas-spin,Aad:2015mxa,cms-spin,cms-higgsprop}. 

In the SM, the \Zg decay proceeds through loop diagrams similar to the \htogg decay. 
The branching ratio for
the Higgs boson decay to $Z\gamma$ is predicted by the SM to be $\brhzg = (1.54\pm0.09)\times 10^{-3}$ at $\mH = 125.09$~\GeV,
which is
comparable with the branching ratio of the Higgs boson decay to a photon pair, $\brhgg = (2.27 \pm 0.05)\times 10^{-3}$. However, the number of reconstructed events is significantly smaller if 
$Z$ boson
decays into electron or muon pairs are considered ($B(Z\to ee)=(3.363\pm 0.004)\%$
and $B(Z\to\mu\mu)=(3.366\pm 0.007)\%$~\cite{Olive:2016xmw}).
A $H\to Z\gamma$ branching ratio different from the SM prediction would be expected if $H$ were
a neutral scalar of different
origin~\cite{low, low2}, or a composite state~\cite{azatov}, or in
models with additional colourless charged scalars, leptons or
vector bosons coupled to the Higgs boson and exchanged in the
\htoZg~loop~\cite{extensions, Chen:2013vi, carena}.

The $H\to Z(\to\ell\ell)\gamma$ decay ($\ell= e$ or $\mu$) 
has been searched for by the ATLAS and CMS
collaborations using the full data sets collected at $\sqrt{s}=7$ and 
8~\TeV~\cite{atlas-lowmass,cms-lowmass}. 
No significant excess over the expected background was observed. 
The ATLAS Collaboration reported an observed (expected) upper limit at the 95\% 
confidence level (CL) of 11 (9) times the SM prediction for a Higgs boson mass of 
$\mH = 125.5$~\GeV. 
The observed (expected) limit from the 
CMS Collaboration is 9.5 (10) times the SM prediction for a 125~\GeV\ Higgs boson.

Many models of physics beyond the SM introduce new heavy bosons $(X)$ through extensions of the Higgs
sector or as additional gauge fields. 
Searches for heavy $Z\gamma$ resonances were carried out by the D0 Collaboration at the 
Tevatron and by the ATLAS and CMS collaborations. 
The D0 Collaboration set upper limits on $\sigma(p\bar{p}\to X) \cdot B(X\to Z\gamma)$ ranging from 
2.5 (3.1)~pb  for a scalar (vector) $X$ mass of 140~\GeV\ to 0.19 (0.20)~pb for a mass of 
600~\GeV~\cite{d0-highmass-2}
using about 1~fb$^{-1}$ of $p\bar{p}$ collision data recorded
at $\sqrt{s} = 1.96$~\TeV.
The ATLAS Collaboration excluded technimesons decaying to $Z\gamma$ for
technimeson masses between 200 and 700~\GeV\ and between 750 and 890~\GeV, and composite spin-0
resonances decaying to $Z\gamma$
for most of the resonance mass range between 
200~\GeV\ and 1.18~\TeV\ for certain model parameters using $pp$ collision data recorded at
\sqes~\cite{atlas-highmass2}. Using 3.2~\invfb\ of $pp$ collision data recorded at \sqtt, the
ATLAS Collaboration set limits on $\sigxbr$ between 295~fb and 8.2~fb
for spin-0 resonances produced in gluon--gluon fusion for a $X$ mass range from 250~\GeV\ to 
2.75~\TeV\ using leptonic and hadronic $Z$ boson decays~\cite{Aaboud:2016trl}. 
The CMS Collaboration set upper limits on $\sigxbr$ between about 300~fb
and about 2.5~fb for spin-0 resonances produced in gluon--gluon fusion for a mass range of 200~\GeV\ to
3~\TeV\ using leptonic and hadronic $Z$ decays and $pp$ collision data taken at $\sqrt{s}=8$ and 
13~\TeV~\cite{cms-highmass2}.

This article presents improved searches for decays of the Higgs boson to $Z\gamma$ as well as 
for narrow high-mass resonances decaying to $Z\gamma$
using $Z$ boson decays to electron or muon pairs. 
The $Z(\to\ell\ell)\gamma$ final state can be reconstructed completely and with high
efficiency, good invariant mass resolution, and relatively small backgrounds. 
It is therefore a powerful experimental signature.
The main background is the non-resonant production of $Z$ bosons in 
conjunction with
photons, which is modelled using samples of simulated events. A smaller contribution 
arises from $Z$ boson production together with hadronic jets when a jet is misidentified
as a photon. This background is studied with a dedicated selection of the photon candidate.  
Both searches are based on $pp$ collision data recorded at
\sqtt~with the ATLAS detector at the LHC during 2015 and 2016, corresponding to a total
integrated luminosity of \lumithirteentev~\invfb.
The search for decays of the Higgs boson to $Z\gamma$ benefits from the increased
Higgs production cross section due to the increased centre-of-mass energy and also from the
larger data set compared to the previous search~\cite{atlas-lowmass}. In addition, the
event categorisation now includes a category sensitive to Higgs boson production 
via vector-boson fusion. The search is optimised based on the expected production
processes and kinematics for a SM Higgs boson.
The search for high-mass resonances uses a 
significantly larger data set than the previous search~\cite{Aaboud:2016trl}
and is extended to spin-2 resonance production.

\section{ATLAS detector and data sample}
\label{sec:samples}
The ATLAS detector~\cite{ATLASDET-2008} at the LHC is a multipurpose particle detector with a forward-backward 
symmetric cylindrical geometry and a near 4$\pi$ coverage in solid angle.\footnote{The ATLAS experiment uses a 
right-handed coordinate system with its origin at the nominal interaction point (IP) in the centre 
of the detector and the $z$-axis along the beam pipe. The $x$-axis points from the IP to the centre of the 
LHC ring, and the $y$-axis points upward. Cylindrical coordinates $(r,\phi)$ are used in the transverse plane, 
$\phi$ being the azimuthal angle around the $z$-axis. The pseudorapidity is defined in terms of the polar 
angle $\theta$ as $\eta=-\ln\tan(\theta/2)$. The transverse energy is defined as $\et = E \sin(\theta)$.}
It consists of an inner tracking
detector, electromagnetic and hadronic calorimeters, and a muon spectrometer.
The inner detector (ID), immersed in a 2 T axial magnetic field provided by a thin superconducting
solenoid, includes silicon pixel and microstrip detectors,
which provide precision tracking in the pseudorapidity range  $|\eta| <2.5$, and a transition-radiation
tracker (TRT) providing additional tracking and information for electron 
identification for  $|\eta| <2.0$. For the \sqtt\
data-taking period, the ID was upgraded with a silicon-pixel insertable B-layer~\cite{ATLASDET-IBL},
providing additional tracking information from a new layer closest to the interaction point.
The solenoid is surrounded by sampling calorimeters: a lead/liquid-argon (LAr) 
electromagnetic calorimeter covering the region $|\eta| <3.2$, a hadronic calorimeter with a
steel/scintillator-tile barrel section for $|\eta| < 1.7$ and two copper/LAr endcaps
for $1.5 < |\eta| < 3.2$. 
The forward region is covered by additional coarser-granularity LAr calorimeters up to $|\eta| = 4.9$. 
The calorimeter is surrounded by the muon spectrometer (MS) consisting of three large 
superconducting toroids each containing eight coils.
Precise momentum measurements for muons with pseudorapidity up to $|\eta| = 2.7$
are provided by three layers of tracking chambers.
The muon spectrometer also includes separate 
trigger chambers covering $|\eta|$ up to 2.4.

A two-level trigger system~\cite{ATLAS:trigperf} was used during the \sqtt~data-taking period. 
The first-level trigger (L1) is implemented
in hardware and uses a subset of the detector information. This is followed by a software-based
level which runs algorithms similar to the offline reconstruction software, reducing the event rate to 
approximately 1~kHz from the maximum L1 rate of 100 kHz.

The $pp$ data collected by ATLAS in 2015 and 2016 were taken at a centre-of-mass energy of $\sqrt{s}$ = 13~\TeV\
and with a bunch spacing of 25~ns. After requiring that the full detector was operational 
during data-taking and application of requirements on the
data quality, the integrated luminosity corresponds to \lumithirteentev~\invfb, of which \intLumiFifteen~and 
\intLumiSixteen~\ifb\ were taken during 2015 and 2016, respectively. 
The average number of $pp$ interactions per bunch crossing (pile-up) ranged from about 13 in 2015 to about 25 in 2016,
with a peak instantaneous luminosity of \peakLumi\ achieved in 2016.

The events were collected with triggers requiring either one or two electrons or muons in the event.
The single-muon trigger has a transverse momentum (\pt) threshold of 26~\GeV\ and applies a 
requirement on the muon track isolation. The track isolation
is defined as the scalar sum of the transverse momenta of the ID tracks 
in a cone of $\Delta R = \sqrt{(\Delta\eta)^2 + (\Delta\phi)^2} = 0.3$ around the muon.
For the trigger used during 2016, the cone size was modified to be $\Delta R = 10 / (\pt/\GeV)$ for
muons with $\pt > 33.3~\GeV$.
The track isolation is computed from ID tracks with
$\pt > 1~\GeV$ and with a longitudinal impact parameter $z_0$ within 6~mm of the $z_0$ of the muon track,
excluding the muon track itself.
The track isolation is required to be less than 
6\% (7\%) of the muon's transverse momentum in the 2015 (2016) data set.
A second single-muon trigger with a \pt\ threshold of 50~\GeV\ has no requirement on the track isolation.
The dimuon trigger has \pt\ thresholds of 22~\GeV\ and 8~\GeV\ 
and does not apply track isolation criteria. Single-electron triggers with 
\pt\ thresholds at 24~\GeV\ (26~\GeV), 60~\GeV, and 120~\GeV\ (140~\GeV) are used, 
as well as a dielectron trigger with
a \pt\ threshold of 12~\GeV\ (17~\GeV) during the 2015 (2016) data taking. 
In the 2016 data-taking period, the lowest-threshold single-electron
trigger required the track isolation in a cone of $\Delta R = 0.2$ for $\pt < 50$~\GeV\ and 
$\Delta R = 10 / (\pt/\GeV)$ for $\pt > 50$~\GeV\ to be less than 10\% of the electron's 
transverse energy.
For all electron triggers, electron candidates must satisfy
identification criteria based on the properties of the energy cluster in the 
electromagnetic calorimeter and its associated track. The single-electron triggers with lower
thresholds use tighter criteria for the electron identification.
For $H\to Z\gamma$ events that pass the analysis preselection (see Section~\ref{sec:eventSelection}), 
the efficiency to pass the trigger selection is 92.9\%~(96.9\%) for $Z$ boson decays to
muon (electron) pairs.
For a high-mass resonance at 1~\TeV,
the corresponding efficiencies are 94.3\% and 99.8\% for muon and electron final states, respectively.

\section{Simulation samples}
\label{sec:mcsamples}
Samples of simulated Monte Carlo (MC) events are used to optimise the search strategy, 
evaluate the selection
efficiency and to study the different background contributions. The 
generated event samples were processed with the detailed ATLAS detector simulation~\cite{ATLAS-SIM} 
based on \textsc{Geant4}~\cite{Geant4} (one exception
is noted below).

For the $H\to Z\gamma$ search, the mass
of the Higgs boson is chosen to be $\mh = 125$~\GeV\ and the corresponding width is 
$\Gamma_H = 4.1$~\MeV~\cite{deFlorian:2016spz}.
The SM Higgs boson production was simulated with 
\textsc{Powheg Box} v2~\cite{Nason:2004rx, Frixione:2007vw, Alioli:2010xd}
using the combined parton distribution function (PDF) set PDF4LHC following the
recommendations~\cite{Butterworth:2015oua} based on the CT14~\cite{Dulat:2015mca}, 
MMHT14~\cite{Harland-Lang:2014zoa} and NNPDF3.0~\cite{Ball:2014uwa} PDF sets and the
Hessian reduction method~\cite{Gao:2013bia, Carrazza:2015aoa, Watt:2012tq}.
The techniques used
for the simulation of gluon--gluon fusion (ggF) production, vector-boson fusion (VBF) production, and
production in association with a vector boson  ($WH$ and $ZH$, together referred to as $VH$) 
and the perturbative order achieved
are summarised in Table~\ref{tab:samples} and detailed below. 
Higgs boson production in association with a $t\bar{t}$ pair and other production
processes are not considered as their contributions to the total Higgs production cross section
are of the order of $0.1\%$ or less.

\begin{table}[htbp]
\caption{Higgs boson production processes produced with \textsc{Powheg Box} with the techniques
used and their precision in $\alpha_\mathrm{s}$ for the event generation (gen.). The total cross section
is known with higher precision in QCD and electroweak (norm.) than available in the event 
generation. The events
were reweighted to reproduce the more precise total cross section.}
\label{tab:samples}
\begin{center}
\begin{tabular}{lllll}
\hline\hline
Process & Technique        & QCD (gen.) & QCD (norm.) & EW (norm.)\\
\hline
ggF     & MiNLO \& NNLOPS & NNLO (incl.), NLO ($H + 1$-jet) & NNNLO      & NLO \\
VBF     & \textsc{Powheg}            & NLO                      & approx. NNLO & NLO \\
$VH$    & MiNLO            & NLO (incl. and $H + 1$-jet) & NNLO         & NLO \\
\hline\hline
\end{tabular}
\end{center}
\end{table}
 
Higgs boson production via ggF was simulated with \textsc{Powheg Box}, using the MiNLO
approach~\cite{Hamilton:2015nsa}, which achieves next-to-leading-order (NLO) precision for both 
the inclusive and the $H + 1$-jet process in quantum chromodynamics (QCD).
In addition, the NNLOPS approach~\cite{Hamilton:2013fea} was used to improve the precision for
inclusive observables to next-to-next-to-leading-order (NNLO) in QCD: 
the Higgs transverse momentum spectrum achieved by 
this technique was found to be in agreement 
with the result obtained using QCD  
resummation with next-to-next-to-leading logarithmic (NNLL)+NNLO precision 
from the HqT calculation~\cite{Bozzi:2005wk, deFlorian:2011xf}. Top
and bottom quark mass effects are included up to NLO precision in QCD. The central scale
choice for the nominal factorisation ($\mu_\mathrm{F}$) and 
renormalisation scales ($\mu_\mathrm{R}$) is $\mu_\mathrm{F} = \mu_\mathrm{R} = m_H/2$. 
The events were reweighted to reproduce the inclusive cross section
at next-to-next-to-next-to-leading-order (NNNLO) precision in QCD and NLO precision in
electroweak corrections~\cite{Anastasiou:2015ema, Anastasiou:2016cez, Actis:2008ug, Anastasiou:2008tj, Butterworth:2015oua, deFlorian:2016spz}.

Higgs boson production via VBF was simulated with \textsc{Powheg Box} at NLO precision in QCD~\cite{Nason:2009ai}. 
The events were reweighted to reproduce the inclusive cross section with approximate-NNLO precision in QCD and NLO 
precision in electroweak corrections~\cite{Butterworth:2015oua, deFlorian:2016spz, Ciccolini:2007jr, Ciccolini:2007ec, Bolzoni:2010xr}. 

Higgs boson production in association with a vector boson via quark--antiquark initial states
was simulated at 
NLO precision in QCD for inclusive events and $H + 1$-jet events using the MiNLO technique~\cite{Hamilton:2012rf}. 
The events were reweighted to reproduce 
the total $VH$ production cross section, including also production via gluon--gluon initial states, at NNLO precision in QCD with NLO electroweak 
corrections~\cite{Butterworth:2015oua, deFlorian:2016spz, Brein:2003wg, Altenkamp:2012sx, Denner:2011id}.

The effects of parton showering, hadronisation and multiple parton interactions (MPI) were
simulated using \textsc{Pythia 8.186}~\cite{Pythia8} configured with the AZNLO set of 
parameters~\cite{Aad:2014xaa} and the CTEQ6L1~\cite{cteq6l1} PDF set.
The events were reweighted to 
reproduce the \htoZg~branching ratio calculated with 
\textsc{Hdecay}~\cite{deFlorian:2016spz, Djouadi:1997yw, Djouadi:2006bz}.

Three additional event samples of gluon--gluon fusion production are used for the studies of theoretical 
uncertainties. 
The first is an event sample generated with \textsc{MadGraph5\_aMC@NLO} version 5.2.3.3 with 
FxFx multijet merging~\cite{Alwall:2014hca,Frederix:2012ps} of
$H + 0$-jet and $H + 1$-jet at NLO precision in QCD, using the NNPDF 3.0 PDF set. 
The decay of the Higgs boson and the 
parton showering, hadronisation and MPI were provided by 
\textsc{Pythia 8.186} using the A14 set of parameters~\cite{ATLASA14} and the 
NNPDF2.3 PDF set~\cite{Ball:2012cx}. 
Two more samples were simulated with
\textsc{Powheg Box} v1~\cite{Nason:2004rx, Frixione:2007vw, Alioli:2010xd, Bagnaschi:2011tu}
 with the CT10 PDF set~\cite{ct10} and 
\textsc{Pythia 8.186} for parton showering and hadronisation using the AZNLO set of parameters and the
CTEQ6L1 PDF set. The two samples were produced with and
 without MPI to study the uncertainties in the signal acceptance related to the
modelling of non-perturbative effects.

Production of $CP$-even, high-mass spin-0 resonances $X$ in the mass range $m_X \in $ [300--2500]~\GeV\ 
was simulated for 
the gluon--gluon fusion and vector-boson fusion production processes and for an 
intrinsic resonance width of 4~\MeV,
which is much smaller than the experimental resolution (see Section~\ref{sec:signalBkgModel})
and referred to as narrow width assumption (NWA). 
Due to the assumed narrow width of
the resonance, the interference between the resonant signal and the non-resonant background
is neglected.
The ggF (VBF) process was simulated for $m_X = 300$, 500, 700, 750, 800, 1000, 1500, 2000 and 2500~\GeV\ 
($m_X =$ 300, 500, 1000 and 2500~\GeV). Both the ggF and VBF processes were 
produced with \textsc{Powheg Box} v1 with the CT10 PDF set.

Production of $CP$-even, high-mass spin-2 resonances $X$ with mass 
$m_X = 250$, 300, 500, 750, 1000, 1500, 2000
and 2500~\GeV\ for an intrinsic resonance width of 4~\MeV\ via gluon--gluon and
quark--antiquark initial states was simulated at LO in QCD in the Higgs Characterisation 
Model~\cite{Artoisenet:2013puc} 
with \textsc{MadGraph5\_aMC@NLO} 2.3.3~\cite{Alwall:2014hca}. 

For the high-mass spin-0 (spin-2) resonances, the parton showering,
hadronisation and MPI were simulated with \textsc{Pythia 8.186} using the AZNLO 
(A14) set of 
parameters and the CTEQ6L1 (NNPDF2.3) PDF set.

The signal shape and the reconstruction and selection efficiency of the studied high-mass
resonances are parameterised as a function of $m_X$. The parameterisation allows
the extraction of the signal shape and efficiency for any mass point at which no simulation sample is available.

The background mainly originates from non-resonant production of a a $Z$ boson and a prompt
photon ($Z$$+$$\gamma$), with a smaller contribution from production of $Z$
bosons in association with jets ($Z+$jets), with one jet misidentified as
a photon. $Z$$+$$\gamma$ production within the SM is primarily due to 
radiation of photons from final-state leptons (FSR) or initial-state quarks (ISR).
Both SM processes were simulated using the \sherpa
generator ~\cite{Gleisberg:2008ta} (version 2.1.1 for $Z$$+$$\gamma$ and version 2.2.0 for 
$Z$+jets), 
and the matrix elements were calculated using the \textsc{Comix}~\cite{Gleisberg:2008fv} and
\textsc{OpenLoops}~\cite{Cascioli:2011va} generators,
where $Z$$+$$\gamma$ production was calculated for real emission of up to two partons
at leading order (LO) in QCD and merged with the \sherpa parton
shower~\cite{Schumann:2007mg} using the ME+PS@LO
prescription~\cite{Hoeche:2009rj}. 
The process of $Z$+jets was calculated for up 
to two partons at next-to-leading-order (NLO)
and four partons at LO and merged with the parton shower using
the ME+PS@NLO prescription~\cite{Hoeche:2012yf}.
For the $Z$$+$$\gamma$ ($Z$+jets) samples, the CT10 (NNPDF3.0) PDF set was used in conjunction
with dedicated parton shower tuning developed by the \sherpa authors.
To study the background model in detail, a large sample of $Z$$+$$\gamma$ events was simulated 
using fast
simulation of the calorimeter response~\cite{fastsim}.

For all event samples, the additional inelastic $pp$ collisions per bunch crossing were simulated
with \textsc{Pythia 8.186}
using the \textit{A2} set of tuned parameters~\cite{Pythia8Tune} and the MRSTW2008LO PDF 
set~\cite{MRST}.
The MC events were reweighted to reproduce
the distribution of the average number of interactions per bunch crossing observed in the data.

Corrections derived from trigger, identification, reconstruction, and isolation efficiency
measurements for electrons and muons, from identification and isolation efficiency measurements
for photons, and from selection efficiency measurements for jets 
are applied to the simulated events to improve the description of the data.
Similarly, energy scale and resolution corrections for all simulated objects are also taken into 
account.

\section{Event selection and categorisation}
\label{sec:eventSelection}
\subsection{Event preselection}

Events are required to have at least one primary vertex candidate, determined using the tracks
with transverse momentum $\pt > 400$~\MeV\ 
reconstructed in the ID. The primary vertex candidate with the largest sum of the
squared transverse momenta of the associated tracks ($\sum\pt^{2}$) is considered to be the primary
vertex of the interaction of interest. For $H/X\to Z\gamma$ signal events, the selected primary vertex 
is within 0.3~mm of the true primary interaction vertex for more than 99\% of the events.

The $H/X\to Z(\to\ell\ell)\gamma$ candidate events are selected by requiring two same-flavour 
opposite-charge leptons 
($\ell = e, \mu$) to form a $Z$ boson candidate and at least one photon candidate.

Muon candidates with $|\eta|<2.5$ are reconstructed by combining tracks in the ID with tracks in 
the MS. To extend the acceptance beyond that of the ID, muon candidates are
reconstructed from tracks reconstructed only in the MS up to $|\eta| = 2.7$~\cite{Aad:2016jkr}. 
Muon candidates are required to satisfy the \textit{medium} criterion and have $\pt > 10$~\GeV.
In order to ensure good track quality, the ID tracks associated with muons in $|\eta|<2.5$ are 
required to have at least one hit in the silicon pixel detector and at least five hits in the silicon 
microstrip detector, as well as to extend into the TRT for $0.1<|\eta|<1.9$.
The muon candidates in
$2.5 < |\eta| <2.7$ are required to have hits
in each of the three layers of MS tracking chambers.

Electron candidates are reconstructed from a cluster of energy deposits in neighbouring cells of
the EM calorimeter and a track, matched to the cluster, in the ID. 
They are required to have $\pt > 10$~\GeV\ and be within the fiducial region $|\eta|<2.47$
excluding the candidates in the transition region between the barrel and endcap EM calorimeters,
$1.37 < |\eta| < 1.52$. The electrons are identified with the \textit{medium} likelihood-based 
criterion~\cite{ATLAS-electrons} built from the shower
shapes of the clusters, the number of hits associated with the track in the ID and the quality of the track-cluster 
matching. 

Both the muon and electron candidates are required to be associated with the primary vertex
by requiring the longitudinal impact parameter, $\Delta\zzero$, computed with respect to the
primary vertex position along the beam-line,
to satisfy $\zzsint < 0.5$~mm, where $\theta$ is the polar angle of the track.
In addition the significance of the
transverse impact parameter $\dzero$ calculated with respect to the measured beam-line position must satisfy
$|\dzero| / \sigma_{\dzero} < 3~(5)$ for muons (electrons) where $\sigma_{\dzero}$ is the 
uncertainty in $\dzero$. 

The efficiency of the muon identification is higher than 99\% (60\%) for $\pt> 10$~\GeV\ muons with 
$|\eta|>0.1$ ($|\eta|<0.1$)~(similar to Ref.~\cite{Aad:2016jkr}), while
the efficiency of the electron identification ranges from about 80\% for electrons with 
$\pt$ = 10~\GeV\ to higher than 90\% for electrons with $\pt >$ 50~\GeV\
(similar to Ref.~\cite{ATLAS-electrons}). 
The efficiency is typically
about 5\% higher in the barrel region of the detector than in the endcaps.

The lepton candidates are further required to satisfy additional criteria for track isolation, 
which is defined similarly to the track isolation used in the trigger (see Section~\ref{sec:samples}),
but uses a different track selection and 
a different cone size in some cases.
The track isolation is computed
as the scalar sum of the transverse momenta of all tracks in a cone around the 
lepton candidate with $\pt > 1$~\GeV\ which satisfy loose track-quality criteria and originate from the 
selected primary vertex, excluding the track associated with the lepton candidate. For muon candidates, 
the cone size is chosen to be $\Delta R = 0.3$ for $\pt < 33.3$~\GeV\ and 
$\Delta R = 10 / (\pt/\GeV)$ for $\pt > 33.3$~\GeV. For electron candidates,
the cone size is chosen to be $\Delta R = 0.2$ for $\pt < 50$~\GeV\ and 
$\Delta R = 10 / (\pt/\GeV)$ for $\pt > 50$~\GeV. The requirement
on the track isolation is chosen such that it is 99\% efficient over the full lepton $\pt$ range.

An overlap removal procedure is applied to the selected lepton candidates. 
If two electrons share the same track, or the two electron clusters 
satisfy $|\Delta\eta| < 0.075$ and $|\Delta\phi| < 0.125$, then only the highest-$\pt$ electron
is retained. Electron candidates that are within 
$\Delta R = 0.02$ of a selected muon candidate are also discarded.

Photon candidates are reconstructed from energy clusters in the electromagnetic 
calorimeter. 
Clusters matched to a conversion vertex, reconstructed from either two tracks consistent with a 
vertex originating from a photon conversion or one track that does not have any hits in the innermost 
pixel layer and has an electron-like response in the TRT, are reconstructed as converted photon candidates. 
Clusters without any matching track (clusters with a matching track are reconstructed as electrons
as described above) or conversion vertex are reconstructed as unconverted photon 
candidates~\cite{Aaboud:2016yuq}.
Photon candidates are required to have $\pt > 10$~\GeV\ and $|\eta|<1.37$ or 
$1.52<|\eta|<2.37$. 
The identification of photon candidates is based on the lateral and longitudinal shape of the 
electromagnetic shower~\cite{Aaboud:2016yuq,ATLAS-photonid}. A \textit{loose} identification is used for preselection
and for background studies.

In order to suppress the events arising from FSR processes 
and $H\to\ell\ell^*\to\ell\ell\gamma$ decays,
photon candidates within a $\Delta R = 0.3$ cone around a selected
electron or muon candidate are rejected.

The selection
criteria described in the preceding paragraphs
define the event preselection for the leptons and photons included
in the reconstruction of the
invariant mass of the $\ell\ell$ and $\ell\ell\gamma$ systems.

The event categorisation described in Section~\ref{sec:categorisation} used in the 
search for decays of the Higgs boson to $Z\gamma$
makes use of hadronic jets produced in association with the Higgs boson 
candidate.
Jets are reconstructed using the anti-$k_t$ algorithm~\cite{Cacciari:2008gp} with a radius parameter of
0.4 with three-dimensional topological clusters as input~\cite{Aad:2016upy}. Jets are corrected on an event-by-event
basis for soft energy deposits originating from pile-up interactions~\cite{ATLAS-jetpucorr} and 
calibrated using a combination of simulation- and data-driven correction factors accounting
for the non-compensating response of the calorimeter and energy loss in inactive regions~\cite{Aaboud:2017jcu}.
Jets are required to have a transverse momentum larger than 25~\GeV\ and $|\eta|<4.4$. 
To reduce the contamination from jets produced in pile-up interactions, jets with transverse momentum smaller 
than 60~\GeV\ and contained within the inner detector's acceptance ($|\eta|<2.4$) are required to
pass a selection based on the jet vertex tagging algorithm~\cite{ATLAS-JVT}, which is 92\% efficient
for jets originating from the hard interaction. The jet vertex tagging algorithm is 
based on the tracks associated with the jet which are consistent with originating from the selected primary vertex.
Jet--lepton and jet--photon overlap removal is performed where the jet is removed if the lepton or photon is 
within a cone of size $\Delta R$ = 0.2.

\subsection{Reconstruction of $Z$ candidates and $H/X$ candidates and final selection}
\label{sec:ZHXreco}

The $Z$ boson candidates are reconstructed from two same-flavour opposite-sign leptons satisfying 
the preselection criteria and with an invariant mass $m_{\ell\ell}$ larger than 45~\GeV. 
Leptons are required to be consistent with the objects that triggered the 
event. Trigger efficiency turn-on effects are mitigated by transverse momentum requirements on the leptons
that fired the single-lepton or dilepton trigger. If the event was triggered by a single-lepton trigger, the
transverse momentum is required to be at least 27~\GeV\ for the leading lepton, and at
least 1~\GeV\ higher than the respective trigger threshold in cases where the event was
triggered by one of the higher-threshold triggers. If the event was triggered by a
dilepton trigger, the transverse momentum is required to be 
at least 24~\GeV\ (18~\GeV) for the 
leading muon (electron) and 10~\GeV\ (18~\GeV) for the subleading muon (electron).
For $Z\to\mu\mu$ candidates with an invariant mass between 66 and 89~\GeV, 
the invariant mass resolution of the $Z$ boson candidate
is improved by correcting the muon momenta for collinear FSR by including 
any reconstructed electromagnetic cluster with \pt above 1.5~\GeV\ lying close
to a muon track (with $\Delta R < 0.15$) if the corrected invariant mass is below 
100~\GeV~\cite{Aad:2014aba}.
A constrained kinematic fit is applied to recompute the 
four-momenta of the dilepton pair~\cite{Aad:2014eva} for $Z\to\mu\mu$ and $Z\to ee$ candidates. 
The fit models the lepton energy and momentum
response as a Gaussian distribution for each lepton, and the Gaussian width is given by the expected resolution.
The $Z$ lineshape is used as a constraint with the approximation of the leptons being massless. The
lineshape is modelled by a Breit--Wigner distribution.
After the application of the FSR corrections and the kinematic fit, $Z$ boson candidates are required to have 
an invariant mass within 15~\GeV\
of the $Z$ boson mass, $m_Z = 91.2~\GeV$~\cite{Olive:2016xmw}. If multiple $Z$ boson candidates pass all
requirements, the candidate with the mass closest to the $Z$ boson mass is chosen.
About 0.2\% (0.5\%) of events that pass the final $H\to Z\gamma$ ($X\to Z\gamma$) selection 
have more than one $Z$ boson candidate within the
15~\GeV\ mass window. 

\begin{figure}[b]
\subfigure[]{\includegraphics[width=.50\textwidth]{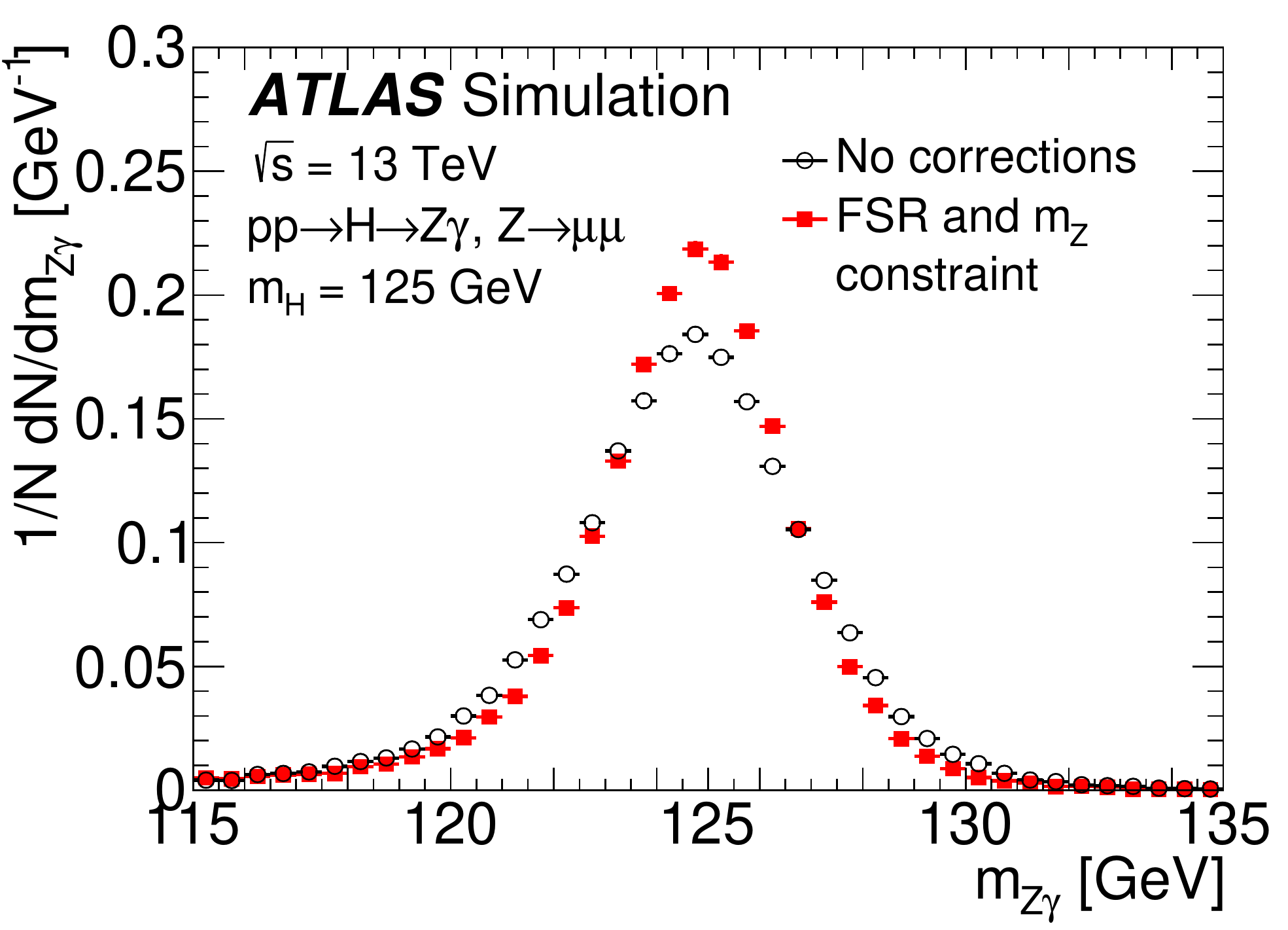}}%
\subfigure[]{\includegraphics[width=.50\textwidth]{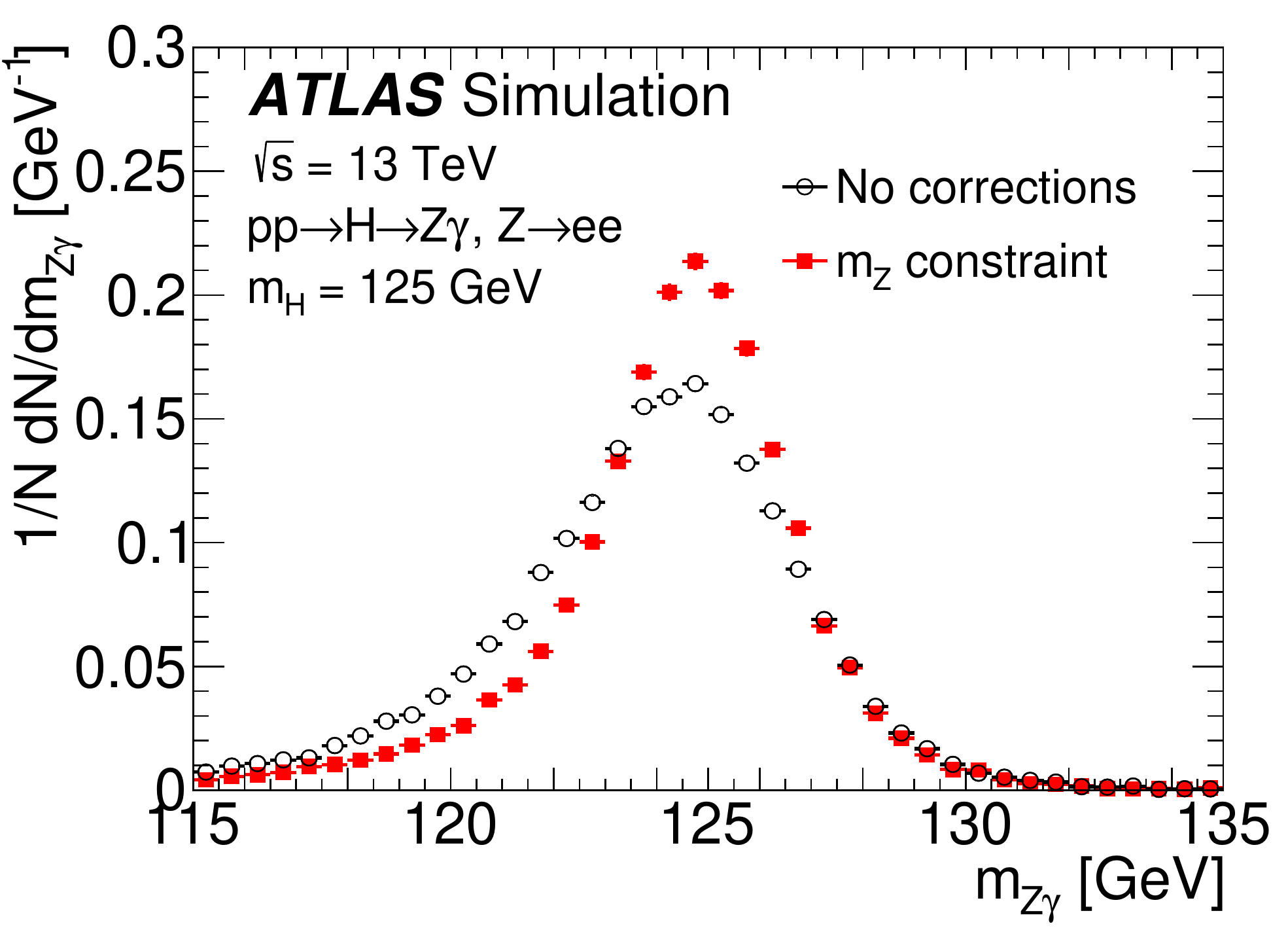}}
\caption{Invariant mass distribution, $m_{Z\gamma}$, for the final selection 
before and after application of the final-state radiation corrections ($Z\to\mu\mu$ only) 
and the $Z$ boson mass 
constrained kinematic 
fit for simulated $H\to Z\gamma$ events with $m_H = 125~\GeV$ in the gluon--gluon fusion production mode. 
Events are separated by lepton type, (a) $Z\to\mu\mu$ and (b) $Z\to ee$.}
\label{fig:mllgamma}
\end{figure}

Higgs boson and $X$ candidates are reconstructed from the $Z$ boson candidate and the highest-$\pt$
photon candidate after the preselection.

For the main analyses with the exception of the background studies, the photon candidate used for the 
reconstruction of the $H/X$ candidate is required to pass the \textit{tight} identification~\cite{Aaboud:2016yuq}. 
The efficiency of the tight identification ranges from 67\% (60\%) to 90\% (95\%)
for unconverted (converted) isolated photons from $\pt$ of 15~\GeV\ to 50~\GeV\ and larger.

Photon candidates are furthermore required 
to be isolated from additional activity in the detector. A combined requirement
on the isolation energy in the calorimeter and the inner detector is used. The calorimeter isolation is
computed as the sum of transverse energies of positive-energy topological 
clusters~\cite{Aad:2016upy} in the calorimeter within a cone of $\Delta R =  0.2$ centred around 
the photon shower barycentre. The transverse energy of the photon
candidate is removed and the contributions of the underlying event and pile-up are subtracted based on
the method suggested in Ref.~\cite{Cacciari:2009dp}. The track isolation for a cone size of 
$\Delta R =0.2$ is used and for converted photons the tracks associated with the
conversion are removed. The calorimeter (track) isolation
is required to be less than 6.5\% (5\%) of the photon \pt.
The efficiency of the isolation requirement for photons satisfying the tight identification criteria 
ranges from approximately 60\% for $\pt$ of 15~\GeV\ to more than 90\% for $\pt$ of 40~\GeV\ 
and larger.

For the $H\to Z\gamma$ ($X\to Z\gamma$) search, the photon transverse momentum requirement is tightened to 15~\GeV\
($\pt / m_{Z\gamma} > 0.3$).
  
The invariant mass of the final-state 
particles, $m_{Z\gamma}$, is 
required to satisfy $115~\GeV < m_{Z\gamma} < 170$~\GeV\ for the $H\to Z\gamma$ search 
and $200~\GeV < m_{Z\gamma} < 2500$~\GeV\ for the high-mass resonance search.
Figure~\ref{fig:mllgamma} shows the invariant 
mass distribution for simulated $H\to Z\gamma$ candidates after the final selection
 with and without the lepton momentum 
corrections from the FSR recovery and the kinematic fit. 
Improvements of the \mmmg\ resolution of 3\% are observed for $m_H=125$~\GeV\ from the FSR
recovery.
The kinematic fit improves the \mmmg\ (\meeg) resolution by 7\% (13\%) at the same mass.
For high invariant masses, the \mmmg\ resolution improvement varies from 10\% at 
$\mX=300~\GeV$ to about 50\% for $\mX > 1.5~\TeV$, while the \meeg\ resolution is
improved by 9\% at $\mX=300~\GeV$ and by 3\% or less above $\mX = 500~\GeV$.
The constrained kinematic fit is particularly effective at large
\mX\ for the $Z\to\mu\mu$ final state due to the decreasing precision of the momentum 
measurement for increasing muon \pt.

\subsection{Categorisation}
\label{sec:categorisation}

Events are split into mutually exclusive event categories that are optimised to improve the sensitivity of 
both the $H\to Z\gamma$ and $X\to Z\gamma$ searches.
The event categories separate events on the basis of the expected signal-to-background
ratio and of the expected three-body invariant mass resolution. Different categories
are used in the search for decays of the Higgs boson to $Z\gamma$ 
and the search for high-mass resonances.

The $H\to Z\gamma$ search uses six exclusive event categories and events are assigned to the categories in
the following order:

\begin{itemize}
\item   \textbf{VBF-enriched} : Events are required to have at least two jets. 
  A boosted decision tree (BDT) that was trained to separate VBF 
  events from other Higgs boson production modes and non-Higgs backgrounds is applied. It uses six kinematic
  variables as input, computed from the $Z$ boson candidate, the photon candidate and the two jets with
  the largest transverse momenta: 
  \begin{itemize}
  \item The invariant mass of the two jets ($m_{jj}$),
  \item The separation of the jets
  in pseudorapidity ($\Delta\eta_{jj}$),
  \item The azimuthal separation of the $Z\gamma$ and the dijet systems
  ($\Delta\phi_{Z\gamma,jj}$),
  \item The component of the transverse momentum of the $Z\gamma$ system 
  that is perpendicular to the difference of the 3-momenta of the $Z$ boson and the photon
  candidate ($\ptt = 2{|p_{x}^{Z} p_{y}^{\gamma} - p_{x}^{\gamma} p_{y}^{Z}|}/{\pT^{Z\gamma}}$),
  \item The smallest $\Delta R$ separation between the $Z$ boson or photon candidate and 
  the two jets ($\Delta R^\mathrm{min}_{Z/\gamma, j}$),
  \item  The difference between the pseudorapidity of the $Z\gamma$ system and the average 
    pseudorapidity of the two jets ($|\eta_{Z\gamma}-(\eta_{j1} + \eta_{j2})/2|$).
  \end{itemize}

  The variable $\ptt$ is strongly correlated with the transverse momentum of the $Z\gamma$ system,
  but has better experimental resolution~\cite{Ackerstaff:1997rc, Vesterinen:2008hx}.
  Any requirement on $\Delta\phi_{Z\gamma,jj}$ effectively vetoes additional jets in the event by 
  restricting the phase space for additional emissions and, to avoid uncontrolled theoretical 
  uncertainties, the BDT does not use shape information for events with $\Delta\phi_{Z\gamma,jj}>2.94$ by
  merging these events into one bin. A minimum value of the BDT output (BDT > 0.82) is required. 
  The expected and
  observed distributions for two input variables, $m_{jj}$ as a typical variable to select
  events with VBF topology and $\Delta\phi_{Z\gamma,jj}$, which serves as an implicit third-jet
  veto, are shown in Figure~\ref{fig:VBFinputs} for
  selected events with at least two jets.
\item \textbf{High relative $\pt$} :
  Events are required to have a high \pt photon, $\pt^\gamma / m_{Z\gamma} > 0.4$.
\item \textbf{$ee$ high $\ptt$} : Events are required to have high $\ptt$ ($\ptt > 40~\GeV$) and a $Z$ boson candidate decay to electrons.
\item \textbf{$ee$ low $\ptt$} : Events are required to have low $\ptt$ ($\ptt < 40~\GeV$) and a $Z$ boson candidate decay to electrons.
\item \textbf{$\mu\mu$ high $\ptt$} : Events are required to have high $\ptt$ ($\ptt > 40~\GeV$) and a $Z$ boson candidate decay to muons.
\item \textbf{$\mu\mu$ low $\ptt$} : Events are required to have low $\ptt$ ($\ptt < 40~\GeV$) and a $Z$ boson candidate decay to muons.
\end{itemize}

\begin{figure}
\subfigure[]{\includegraphics[width=.5\textwidth]{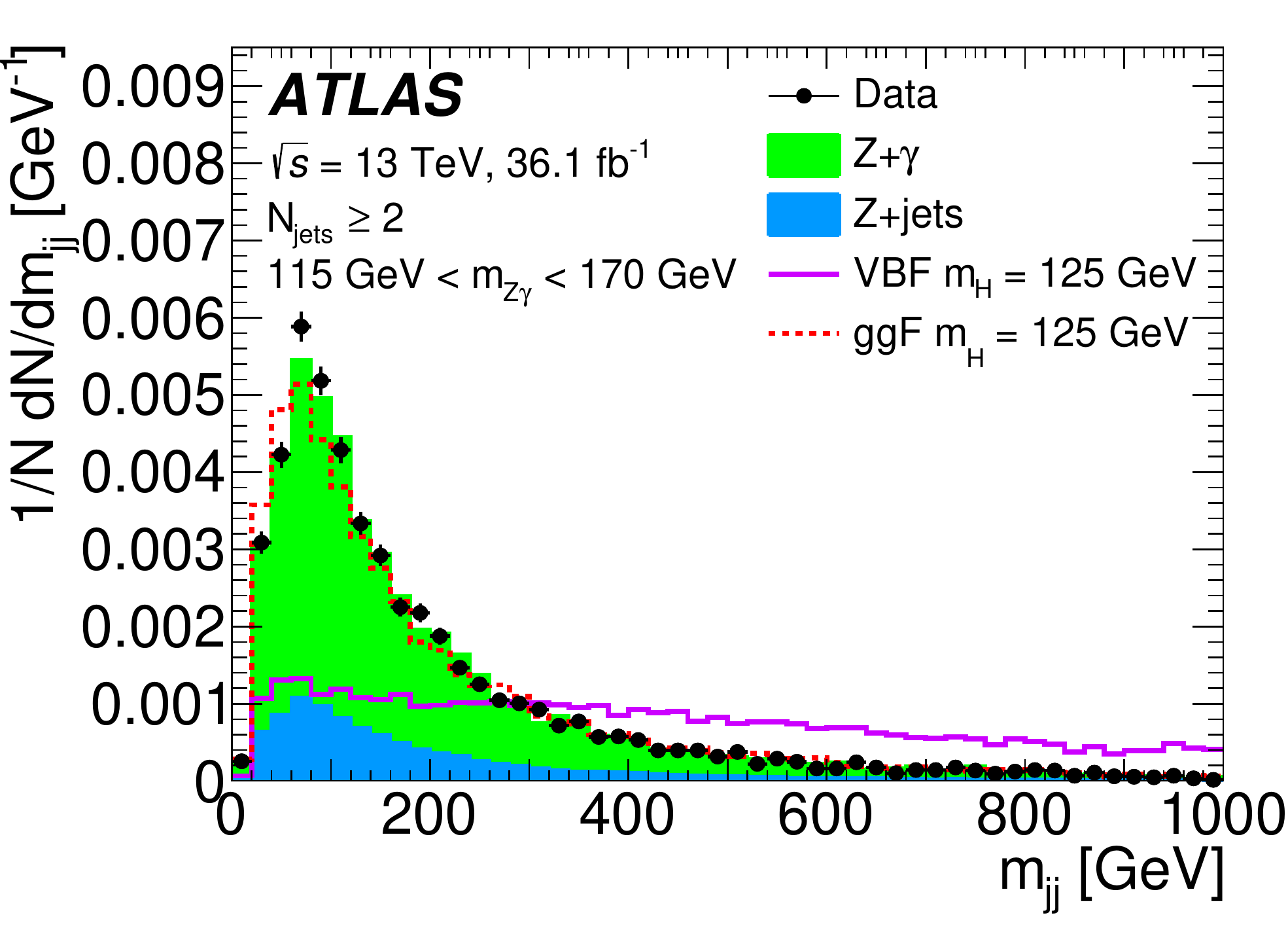}}%
\subfigure[]{\includegraphics[width=.5\textwidth]{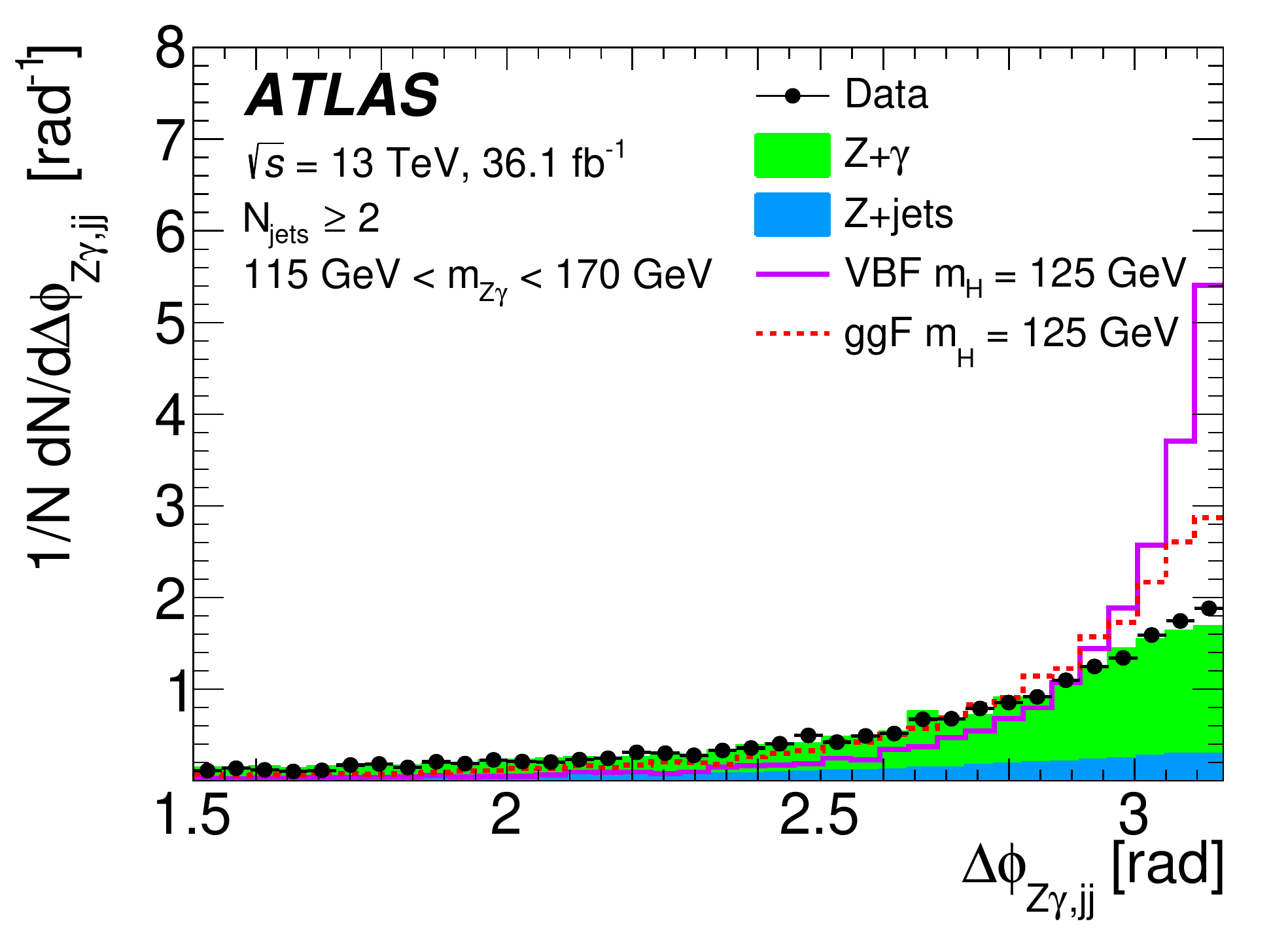}}
\caption{Kinematic variables used in the BDT used to define the VBF-enriched category: 
(a) the invariant mass of the two jets with the highest
transverse momenta, $m_{jj}$ and (b) the azimuthal separation of the $Z\gamma$ and the dijet system,
$\Delta\phi_{Z\gamma, jj}$ for events with at least two jets and 
$115~\GeV < m_{Z\gamma} < 170~\GeV$. 
The observed distribution (normalised to unity) is shown as data points.
The contributions from $Z+\gamma$ events
(obtained from simulation) and the contribution from $Z$+jets 
(obtained from data control regions described in the text) are shown as stacked histograms.
The corresponding expected distributions for Higgs bosons produced via gluon--gluon fusion and vector-boson 
fusion production for $m_H = 125~\GeV$ are shown as open histograms.
The 
$\Delta\phi_{Z\gamma, jj}$ distribution is shown before the suppression of the shape information
for $\Delta\phi_{Z\gamma, jj}>2.94$.}
\label{fig:VBFinputs}
\end{figure}

\begin{table}
\caption{The expected signal efficiency times acceptance, denoted by $\epsilon$, per production mode 
for each category after the 
full event selection, as well as the expected fraction $f$ of each production
process relative to the total signal yield, for simulated
SM Higgs boson production assuming $m_H = 125~\GeV$.
The expected number of signal events per production process is also given.}
\label{tab:cat-eff}
\begin{center}
\begin{tabular}{lSSSSSSSS}
\hline\hline
& \multicolumn{2}{c}{ggF} & \multicolumn{2}{c}{VBF} &\multicolumn{2}{c}{$WH$} & \multicolumn{2}{c}{$ZH$}\\
Category &  $\epsilon$[\%] & $f$[\%] & $\epsilon$[\%] & $f$[\%] & $\epsilon$[\%]  & $f$[\%] & $\epsilon$[\%] & $f$[\%] \\
\hline
VBF-enriched                  & 0.25 & 30.5 & 6.5 & 67.5 & 0.34 & 1.3 & 0.24 & 0.6  \\
High relative $\pt$           & 1.1  & 71.5 & 2.6 & 14.3 & 4.0 & 8.3 & 4.1 & 5.3  \\
$ee$ high $p_{\mathrm{T}t}$     & 1.7 & 80.8 & 2.8 & 11.0 & 3.2 & 4.7 & 3.6 & 3.3  \\
$ee$ low $p_{\mathrm{T}t}$      & 7.1 & 93.2 & 3.6 & 4.1 & 3.7 & 1.5 & 4.2 & 1.1  \\
$\mu\mu$ high $p_{\mathrm{T}t}$   & 2.2 & 80.4 & 3.6 & 11.3 & 4.1 & 4.8 & 4.2 & 3.1  \\
$\mu\mu$ low $p_{\mathrm{T}t}$    & 9.2 & 93.4 & 4.7 & 4.1 & 4.6 & 1.5 & 4.8 & 1.0  \\

\hline                                          
 Total efficiency (\%)   & 21.5 &  & 23.8 &  & 20.2 &  & 21.0 &   \\
\hline                 
Expected events & \multicolumn{2}{c}{35} & \multicolumn{2}{c}{3.3} &\multicolumn{2}{c}{1.0} & \multicolumn{2}{c}{0.7} \\
\hline
\hline
\end{tabular}
\end{center}
\end{table}

For SM $H\to Z(\to\ell\ell)\gamma$ events, the reconstruction and selection efficiency (including
kinematic acceptance) is  21.5\%.
Table~\ref{tab:cat-eff} shows the expected signal efficiency times acceptance for each of the different SM
Higgs boson production processes in each category, as well as the expected relative contribution of
a given production process to each category.
The VBF-enriched category is expected to be about 68\% pure in VBF events. The high relative 
$\pt$
and high $\ptt$ categories are expected to be slightly enriched in VBF and $VH$ events. Overall,
about 40 $H\to Z\gamma$ events are expected to be selected.
Table~\ref{tab:SBtable} summarises for each category: 
the number of selected events from data in the fit range $115~\GeV < m_{Z\gamma} < 150~\GeV$
(see Section~\ref{sec:signalBkgModel}); 
the expected number of events ($S_{90}$) in an interval around the $m_{Z\gamma}$ peak position 
expected to 
contain 90\%  of the SM signal events;
$w_{90}$ defined to be half of the width of the interval; 
the expected $S_{90}/B_{90}$, where $B_{90}$ is the background yield in the same mass window determined from data;
and the expected $S_{90}/\sqrt{S_{90}+B_{90}}$.
The window is constructed so that it includes 45\% of the signal events on either side of the peak position
for $\mH=125.09~\GeV$. The largest fraction of the signal is expected in the low \ptt\ 
categories, which have the smallest expected significance. The VBF-enriched category shows
the largest contribution to the expected significance.

\begin{table}[htbp]
\caption{The number of data events selected in the mass range used for the background fit to the 
$m_{Z\gamma}$ spectrum (115--150~\GeV) per category. In addition, the following numbers are given:
the expected number of Higgs boson signal events in an interval around the peak position for a signal 
of $\mH=125.09~\GeV$, 
expected to contain 90\% of the SM signal ($S_{90}$), the half-width of the $S_{90}$ interval 
($w_{90}$), as well as
the expected signal-to-background ratio in the $S_{90}$ window ($S_{90}/B_{90}$) with 
$B_{90}$ determined from data, and the expected significance estimate $S_{90}/\sqrt{S_{90}+B_{90}}$.}
\label{tab:SBtable}
\begin{center}
\begin{tabular}{lrSccc}
\hline\hline
                 Category  & Events &  $S_{90}$  & $w_{90}$ [GeV] & $S_{90}/B_{90}$ $[10^{-2}$] & $S_{90}/\sqrt{S_{90}+B_{90}}$\\
\hline
VBF-enriched                 & 88   & 1.2  & 3.9 & 9.5 & 0.32 \\        
High relative $\pt$          & 443   & 2.3  & 3.9 & 3.0 & 0.26  \\         
$ee$ high $p_{\mathrm{T}t}$      & 1053  & 3.3  & 3.9 & 1.1 & 0.19 \\         
$ee$ low $p_{\mathrm{T}t}$       & 11707 & 11.2 & 4.2 & 0.3 & 0.18 \\         
$\mu\mu$ high $p_{\mathrm{T}t}$  & 1413  & 4.0  & 3.7 & 1.2 & 0.22 \\         
$\mu\mu$ low $p_{\mathrm{T}t}$   & 16529 & 14.5 & 3.8 & 0.3 & 0.21 \\         
\hline\hline
\end{tabular}
\end{center}
\end{table}

The search for high-mass resonances uses two categories, one for each $Z$ 
boson candidate decay mode, $Z\to ee$
and $Z\to\mu\mu$, to benefit from both the better invariant mass resolution in the electron channel at 
large $m_{Z\gamma}$ and the differences in the systematic uncertainties between electrons and muons.
The invariant mass resolution, measured by the Gaussian width of the signal model (see Section~\ref{sec:signalBkgModel}),
ranges from 2.8~\GeV\ (3.1~\GeV) at $m_X = 250~\GeV$ to 16~\GeV\ (36~\GeV) at $m_X = 2.4~\TeV$
for $Z\to ee$ ($Z\to\mu\mu$).

\section{Signal and background modelling}
\label{sec:signalBkgModel}
The signal and background yields are extracted from the
data $m_{Z\gamma}$ distribution by assuming analytical models.
The parameters that describe the shape of the signal are obtained from simulated signal samples. 
The analytical models used for the background shape are chosen using simulated background
samples and the values of their free parameters are determined from the fit to data.

\subsection{Signal modelling}
\label{sec:signalModel}

The signal mass distribution in the searches for both the Higgs boson and the high-mass resonance decay to $Z\gamma$ is well 
modelled with a double-sided
Crystal Ball (DSCB) function (a Gaussian function with power-law tails on both sides)~\cite{dscb1, dscb2}.
The peak position 
and width
of the Gaussian component are represented by $\mu_\mathrm{CB}$ and $\sigma_\mathrm{CB}$, respectively.

To determine the parameters of the DSCB for the $H\to Z\gamma$ search, a 
fit is performed to all the categories (see Section~\ref{sec:categorisation})
from the simulated signal samples produced via ggF, VBF and $VH$ processes at $\mH = 125~\GeV$. 
A shift of 90~\MeV\ is applied
to the peak position $\mu_\mathrm{CB}$
to build a signal model for $\mH = 125.09~\GeV$.
For the high-mass search, a simultaneous fit is performed to all signal samples,
$\mX = $ [300--2500]~\GeV\ ($\mX = $ [250--2500]~\GeV) for the spin-0 (spin-2) interpretation. 
This allows a parameterisation of
the signal shape for masses \mX\ for which no simulation sample is available.
The \mX\ dependence of the signal shape parameters is parameterised by polynomials, 
and their coefficients are determined during the simultaneous fit.
The parameterisation is done separately for each of the three models considered, a spin-0 resonance
and a spin-2 resonance produced via either gluon--gluon or quark--antiquark initial states.

\begin{figure}
\begin{center}
\subfigure[]{\includegraphics[width=.50\textwidth]{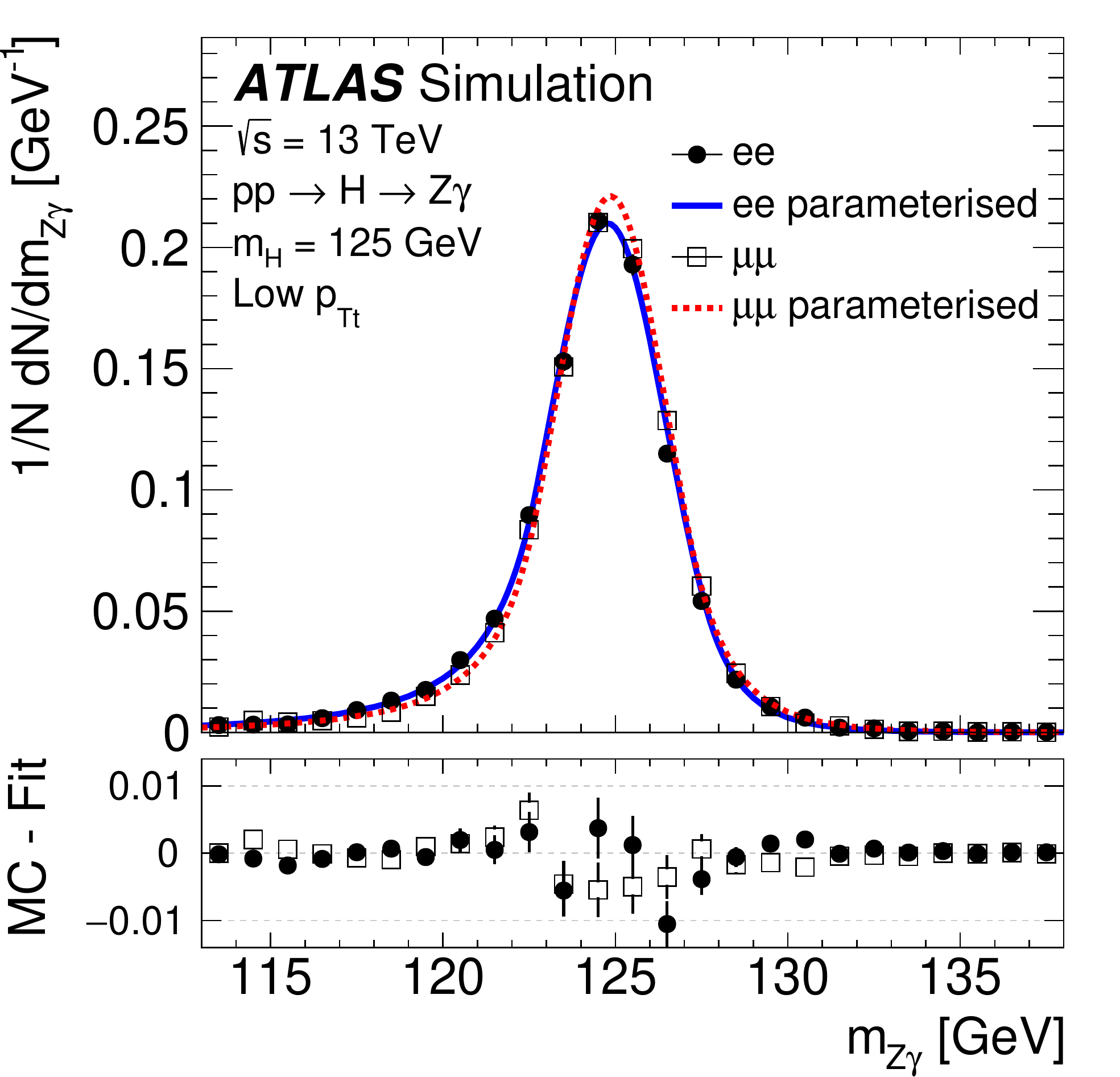}}%
\subfigure[]{\includegraphics[width=.50\textwidth]{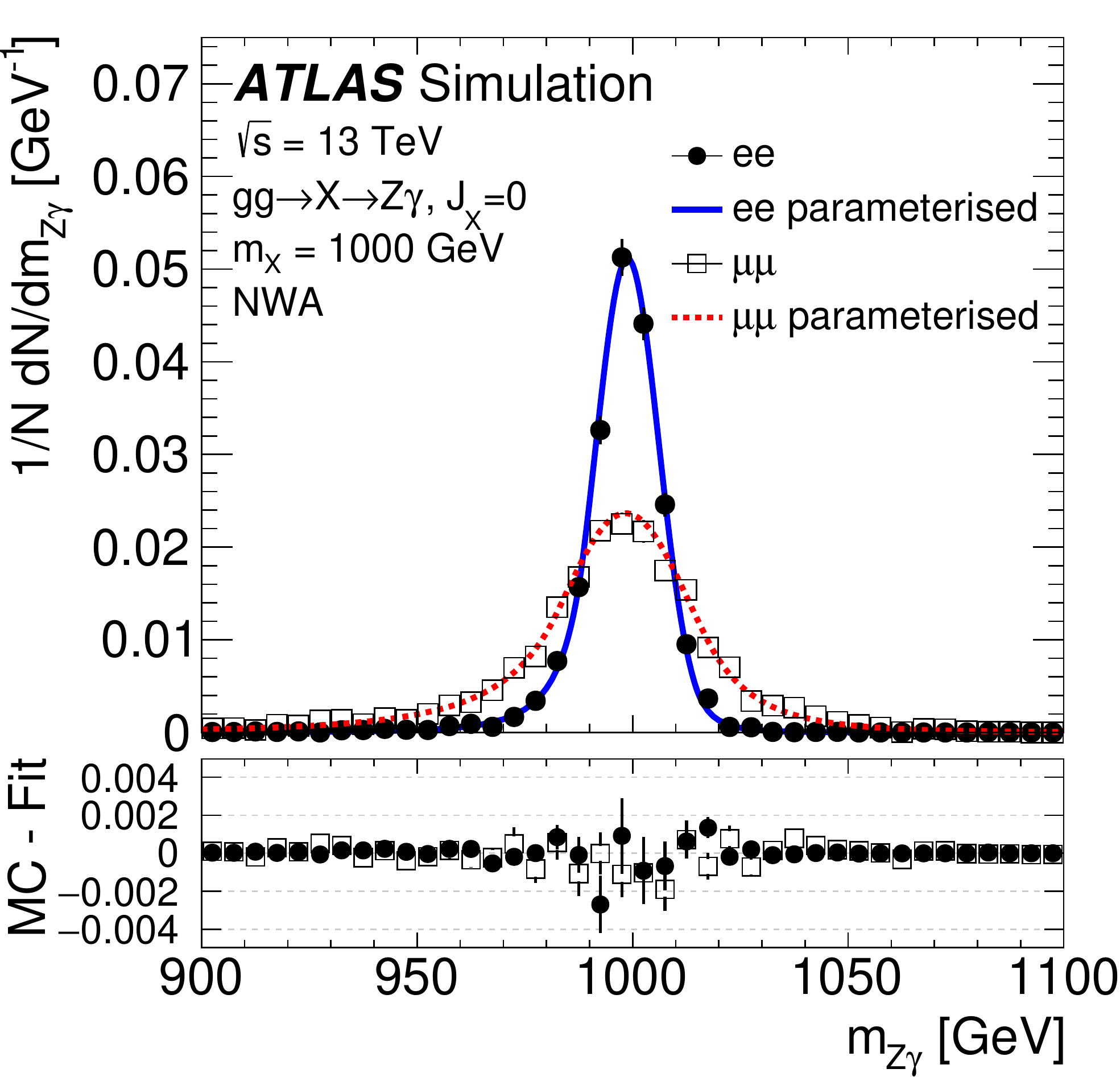}}
\end{center}
\caption{The differential distribution of the invariant $Z\gamma$ mass ($m_{Z\gamma}$) for (a) Higgs
bosons with $m_H=125~\GeV$ in the low \ptt\ 
categories and (b) high-mass spin-0 particles produced 
via gluon--gluon fusion and with $m_X=1000~\GeV$, using the narrow width assumption (NWA). 
The markers show the $m_{Z\gamma}$ distributions and 
the solid and dotted lines the fitted parameterisations used in the searches.
The bottom part of the figures shows the residuals between the markers and the 
parameterisation.}
\label{fig:sig-mmg-fit}
\end{figure}

Figure~\ref{fig:sig-mmg-fit} shows the MC-simulated $m_{Z\gamma}$ distribution at 
$\mH = 125$~\GeV\ for the low \ptt categories and at $\mX = 1000$~\GeV. 
Similar fit qualities are obtained for all the categories in both searches.

Additionally, the signal efficiency defined as the number of events satisfying all the selection 
criteria (as given in Section~\ref{sec:eventSelection}) normalised to the
total number of events is needed to extract \sigbr ~from the measured yield.
For the $H\to Z\gamma$ search,
the signal efficiency times the acceptance in each category are shown in 
Table~\ref{tab:cat-eff}.\footnote{The efficiency difference between $m_H = 125~\GeV$ and
$m_H = 125.09~\GeV$ is estimated to be smaller than 1\%.}  

For the search for high-mass resonances, the signal efficiency is parameterised as a function 
of the resonance mass with an exponentiated second-order polynomial. 
Figure~\ref{fig:efficiency}(a) shows the
reconstruction and selection efficiency for $X\to Z(\to\ell\ell)\gamma$ events for a spin-0 resonance
produced in gluon--gluon fusion, separately for $Z\to ee$ and $Z\to\mu\mu$. 
The efficiencies range from about 30\% to about 46\% in the mass range from
250~\GeV\ to 2.4~\TeV. For a spin-0
resonance produced via vector-boson fusion, the efficiency is larger by up to 4\% over the full
resonance mass range considered.
Figure~\ref{fig:efficiency}(b) shows the reconstruction and selection efficiency for 
spin-2 resonances produced via gluon--gluon and quark--antiquark initial states 
as a function of the resonance mass. 
For spin-2 resonances
produced in gluon--gluon (quark--antiquark) initial states, the efficiencies range from about 22\% (28\%) to about 
35\% (54\%) in the mass range from 250~\GeV\ to 2.4~\TeV.
The efficiency differences between the spin-0 resonance
produced via gluon--gluon fusion, the spin-2 resonance produced via gluon--gluon initial states
and the spin-2 resonance produced via quark--antiquark initial states are primarily related to
the different photon transverse momentum distributions between the different production mechanisms.

\begin{figure}
\begin{center}
\subfigure[]{\includegraphics[width=0.5\textwidth]{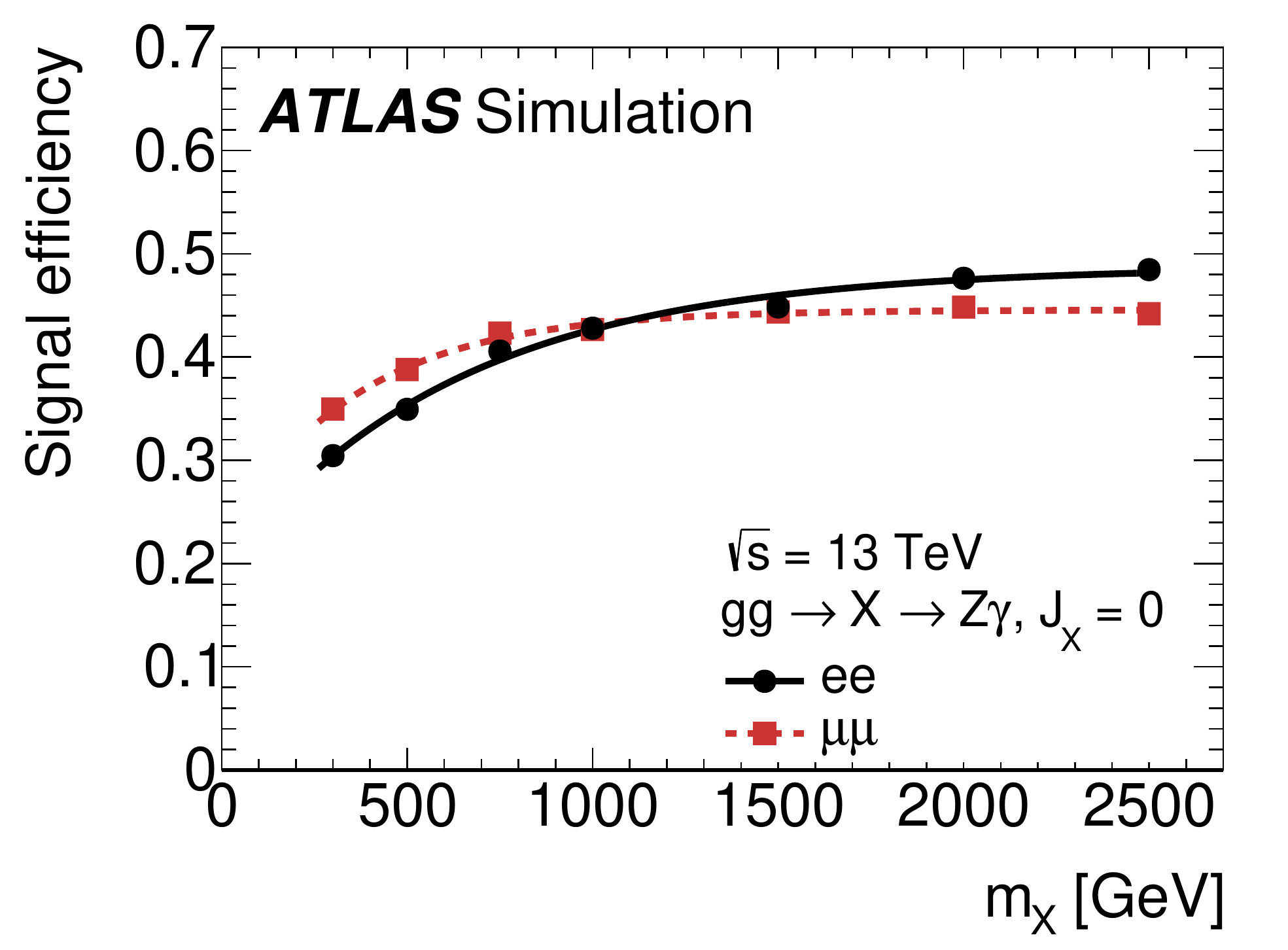}}%
\subfigure[]{\includegraphics[width=0.5\textwidth]{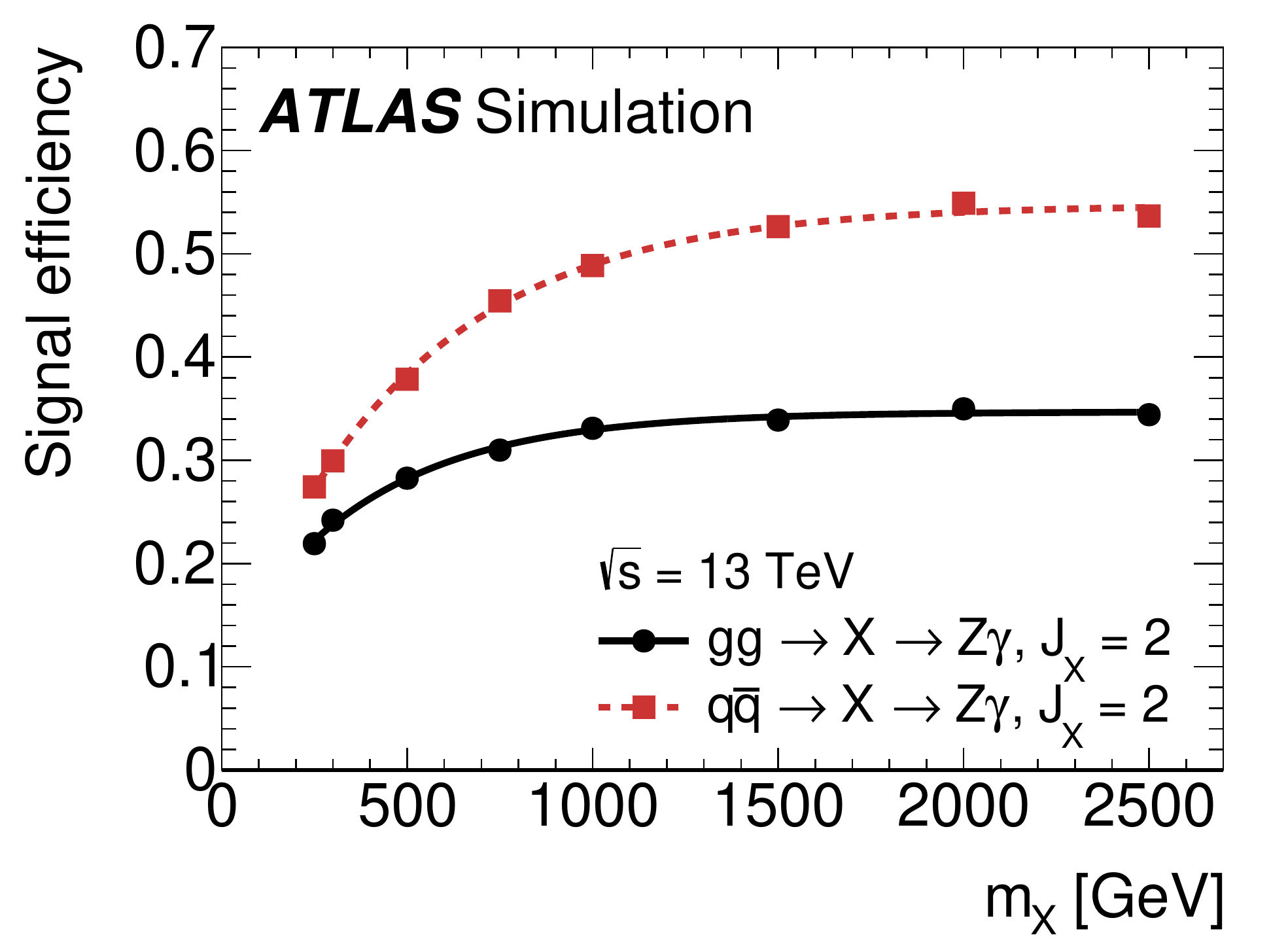}}
\end{center}
\caption{Reconstruction and selection efficiency (including kinematic acceptance) for the 
$X\to Z\gamma$ final state as a function of the resonance mass $m_X$ (a) for
a spin-0 resonance via gluon--gluon fusion, separately for the $ee$ and the 
$\mu\mu$ categories, and 
(b) for a spin-2 resonance produced via either the gluon--gluon or the quark--antiquark 
initial states.
The markers show the efficiencies obtained from simulation, while the curves represent the 
parameterisation used in the analysis. The efficiencies are given with respect to (a)
$X\to Z(\to ee)\gamma$ and $X\to Z(\to \mu\mu)\gamma$, respectively, and (b) 
$X\to Z(\to \ell\ell)\gamma$ where $\ell = e, \mu$.}
\label{fig:efficiency}
\end{figure}

\subsection{Background modelling}
\label{sec:bckgModel}
The background is mainly composed of non-resonant production of a $Z$ boson in association with
a photon (irreducible background), and of inclusive $Z+$jets events where a jet is misidentified as 
a photon (reducible background), and the relative contributions are determined using data as described
below. Contributions from other sources, such as $t\bar{t}$ production, $W/Z$ events, and, 
for the
$H\to Z\gamma$ search, from other Higgs 
boson decays are expected to be negligible based on studies of simulated events. 
The background exhibits a smoothly falling distribution 
as a function of the invariant mass of the candidate $Z$ boson and photon, \mzg. 

The estimated background composition is used to construct simulated background
samples with the same composition as the data background.
These samples are used in the optimisation
of the selection criteria, the choice of analytical model of the
background shape, and the estimation of the related systematic uncertainties.
The searches rely 
only indirectly on the measured background composition 
since the background shape parameters are determined from the data.

The composition of the background is estimated using a combined binned fit to the calorimeter isolation 
distribution of the photon candidate in the signal region and in a control region enriched in
$Z$+jets background. In the control region, 
photon candidates are required to fail the tight identification, but to pass a modified
loose identification. It differs from the tight identification by removing the requirements on four 
out of nine 
shower shape variables which are the least correlated with the calorimeter isolation~\cite{Aad:2010sp}.
The calorimeter isolation distribution for photons and the contribution of true photons to the control 
region are determined from simulation, 
while the calorimeter isolation distribution for jets is determined 
in the fit and assumed to be the same in the signal and 
control regions. This assumption is supported by extensive studies performed in the context
of earlier analyses~\cite{Aad:2010sp}. The composition is determined in the inclusive 
selection for the $H\to Z\gamma$ search and the fraction of $Z$$+$$\gamma$
events is found to be $0.838\pm 0.005\,\mathrm{(stat.)} \pm 0.031\,\mathrm{(syst.)}$. For the 
high-mass resonance search, the fraction of
$Z$$+$$\gamma$ events is found to be $0.916\pm 0.009\,\mathrm{(stat.)}\,^{+0.013}_{-0.019} 
\,\mathrm{(syst.)}$. 
The systematic uncertainties are estimated by
varying the set of shower shape variables that are removed from the tight identification to
define the modified loose identification~\cite{Aad:2010sp}.
The results of the composition estimate are cross-checked with a
two-dimensional sideband technique~\cite{Aad:2010sp} based on the calorimeter isolation of the photon
candidate and whether or not the photon candidate satisfies the tight identification criteria
(when the photon fails the tight identification it is still required to pass the modified
loose identification), which gives consistent results.

The analytical model of the background and the $m_{Z\gamma}$ range used for the final fit are chosen to limit the bias in the extracted signal 
yield, while at the same time limiting the number of free parameters in the fit to avoid degradation
of the sensitivity~\cite{atlas-higgs}. For
each category used in either analysis, the 
bias (also referred to as spurious signal) is estimated as a function of the signal invariant
mass by performing a signal+background fit to a 
$m_{Z\gamma}$ background-only distribution with small statistical fluctuations. 
The background-only distribution is
constructed from the fast simulation of $Z$$+$$\gamma$ events, and the contribution from 
$Z$+jets events is included by reweighting the $Z$$+$$\gamma$ simulated distribution as follows: 
for each category, the shape of the $Z$+jets contribution is
determined in a data control region defined by requiring that the photon candidate fails to satisfy the identification 
and isolation criteria. To smooth statistical fluctuations in the $Z$+jets shape, a first-order
polynomial is fitted to the ratio of the $Z$+jets and $Z$$+$$\gamma$ shapes, and the smoothed $Z+$jets shape 
is constructed from the fit result and the $Z$$+$$\gamma$ shape. The reweighting of the $Z$$+$$\gamma$ 
distribution to take into account the $Z$+jets contribution is determined from the smoothed
$Z$+jets shape. The normalisation of the $Z$+jets contribution is determined from the number of
events obtained when applying the selection and categorisation to the $Z$$+$$\gamma$ and $Z$+jets simulation
samples and the purity of the inclusive sample, measured as described above.
The spurious signal is required to be less than 40\% (20\%) of the expected statistical
uncertainty in the signal yield, which is dominated by the expected statistical uncertainty of 
the background, in the search for Higgs boson (high-mass resonance) decays to 
$Z\gamma$. The looser requirement for the $H\to Z\gamma$ search is chosen to improve the robustness
of the procedure against statistical fluctuations in the simulated $Z$$+$$\gamma$ event sample.
If two or more considered functions satisfy this requirement, the
function with the fewest number of parameters is chosen. 

For the $H\to Z\gamma$ search, the fit range is also optimised on the basis of the spurious signal
estimates, taking into account the spurious signal and the number of parameters of the chosen functions
in all categories. A fit range from $115~\GeV$ to $150~\GeV$ is selected. 
A second-order Bernstein polynomial
is chosen as the parameterisation for the VBF and high relative $\pt$ categories, and a fourth-order
Bernstein polynomial is chosen for the other categories. For the chosen parameterisation, the largest spurious 
signal obtained in a window of 121--129~\GeV\ is
assigned as a systematic uncertainty in each category associated with the choice of background 
function and ranges from
1.7 events in the VBF category to 25 events in the $\mu\mu$ low \ptt\ category.
The choice of background functions is validated by using second- and third-order polynomials for the
smoothing of the $Z+$jets background shape and by varying the $Z$$+$$\gamma$ purity by $\pm 15\%$.
The large variation of the $Z$$+$$\gamma$ purity is chosen to cover the purity differences between
the different categories and intended to also account for the additional uncertainty in the estimation
of the $Z$+jets invariant mass distribution.

The high-mass resonance search considers as a background model a class of functions~\cite{Aaboud:2016trl} given by

\beq
f_\mathrm{bkg}^k(x; b, {a_k}) = N (1-x^{1/3})^{b}x^{\sum_{j=0}^k a_k \log(x)^j},
\label{eqn:background}
\eeq

where $x= m_{Z\gamma}/\sqrt{s}$, $N$ is a normalisation factor, $k$ determines the number of terms considered in the exponent, and
$b$ and $a_k$ are determined by the fit. 
When testing on the background-only distribution constructed using the simulated $Z$$+$$\gamma$ 
sample taking into account $Z$+jets contributions as discussed before, the spurious signal
criterion is found to be satisfied for the full mass range for $k=0$.
The spurious signal used as an estimate of the systematic uncertainty is parameterised as a smooth function
of the invariant mass. It ranges from 3.6 (6.1) events at 250~\GeV\ to 0.01 (0.005) events at 2.4~\TeV\ for the
$Z\to\mu\mu$ ($Z\to ee$) channel.
The choice of analytic function for the background shape is validated by
using second- and third-order polynomials for the
smoothing of the $Z$+jets background shape, by varying the $Z$$+$$\gamma$ purity by $\pm 5\%$ (motivated by the
range of purities estimated in the two categories), and
by varying the PDFs in the $Z$$+$$\gamma$ simulation 
using the uncertainties associated with the different eigenvectors of the PDF set.

The possibility of needing higher-order functions when fitting
the selected analytical function to the data $m_{Z\gamma}$ distribution
is further investigated by an $F$-test. The test statistic $F$ defined as

\beq
F = \left(\frac{\chi^2_0 - \chi^2_1}{p_1-p_0}\right)/\left(\frac{\chi_1^2}{n-p_1}\right),
\eeq

compares the fit qualities between less and more complex functions.
The $\chi^2_0$ ($\chi^2_1$) is the $\chi^2$ value of a binned fit with the less (more) complex parameterisation,
$p_k$ is the number of free parameters of each fit, and $n$ is the number of bins of the invariant 
mass distribution. Should the probability to find values of $F$ more extreme than the one measured on data 
be less than 5\%, the less complex parameterisation would be rejected in favour of the more complex
parameterisation. 
The binning for the $F$-test is chosen to guarantee a sufficient number of events in
each bin.
For the $H\to Z\gamma$ search, the test is carried out to determine if there is
any indication that a higher-order Bernstein polynomial is required. 
It does not lead to a change in the chosen parameterisation.
For the high-mass search, the test is
performed to determine whether or not the quality of the fit to data is improved significantly if 
using $k=1$. 
The test confirms that the choice of $k=0$ is adequate. The $\chi^2$ per degree of freedom is 1.2 for 30
degrees of freedom (1.1 for 17 degrees of freedom) or better for the chosen parameterisations 
for the $H\to Z\gamma$ (high-mass resonance) search.

\section{Systematic uncertainties}
\label{sec:sys}
The experimental and theoretical uncertainties that are considered in the searches can be 
grouped into three classes:
uncertainties associated with the parameterisation of the signal and background distributions
(see Section~\ref{sec:uncshape}), experimental uncertainties in the efficiency and acceptance affecting
the expected event yields (see Section~\ref{sec:uncyield}), and theoretical uncertainties in the 
modelling of
the signal in the simulation (see Section~\ref{sec:unctheo}). The nuisance parameters in the likelihood 
function (see Section~\ref{sec:statistics}) represent the uncertainties which
are studied in each category using the simulated signal samples generated at $\mH =125$~\GeV\ for the 
$H\to Z\gamma$ search, and 
$\mX = [300-2500]$~\GeV\ at high-mass.
The main experimental sources of uncertainty are summarised in Table~\ref{tab:syst}.

\begin{table}
  \caption{The main sources of experimental uncertainty for the $H/X\to Z \gamma$ searches. The gluon--gluon fusion signal samples
produced at $\mh = 125~\GeV$ and
$\mX = $ [300--2500]~\GeV\ are used to estimate the systematic uncertainty.
The ranges for the uncertainties span the variations among different categories and different \mX~resonance masses. 
The uncertainty values are given as fractions of the total predictions, except for the spurious signal uncertainty, 
which is reported as the absolute number of events. Values are not listed if systematic sources are negligible or 
not applicable.
}
\label{tab:syst}
  \centering
  \begin{tabular}{lr@{--}l@{\hspace{1.5ex}}r@{--}l@{}}
    \hline
    \hline       
Sources & \alignUnderEnDash{\htoZg} &  & \alignUnderEnDash{\xtozg} & \\
\hline
     \hline

     \multicolumn{5}{c}{\em Luminosity $\mathrm{[\%]}$}                          \\
     \hline
     Luminosity                         & \alignUnderEnDash{3.2} & & \alignUnderEnDash{3.2} &                 \\
     \hline
    \multicolumn{5}{c}{\em Signal efficiency $\mathrm{[\%]}$}                          \\
     \hline
     Modelling of pile-up interactions    & 0.02 & 0.03 & $<0.01$ & 0.2     \\
     Photon identification efficiency     & 0.7 & 1.7   & 2.0 & 2.6     \\
     Photon isolation efficiency          & 0.07 & 0.4  & 0.6 & 0.6     \\
     Electron identification efficiency   & 0.0 & 1.6  & 0.0 & 2.6     \\
     Electron isolation efficiency        & 0.0 & 0.2  & 0.0 & 3.5     \\
     Electron reconstruction efficiency   & 0.0 & 0.4  & 0.0 & 1.0     \\
     Electron trigger efficiency          & 0.0 & 0.1  & 0.0 & 0.2     \\
     Muon selection efficiency            & 0.0 & 1.6  & 0.0 & 0.7     \\
     Muon trigger efficiency              & 0.0 & 3.5  & 0.0 & 4.2     \\
     MC statistical uncertainty           &  &         & 1.2 & 2.0      \\
     Jet energy scale, resolution, and pile-up   & 0.2 & 10  &  &   \\
    \hline
    Total (signal efficiency)              & 2.1 & 10 & 4.0 & 6.3\\
    \hline
    \multicolumn{5}{c}{\em Signal modelling on $\sigma_\mathrm{CB}$ $\mathrm{[\%]}$} \\
    \hline
    Electron and photon energy scale       &  0.6 & 3.5  & 1.0 & 4.0        \\
    Electron and photon energy resolution  &  1.1 & 4.0  & 4.0 & 30        \\   
    Muon momentum scale                    &  0.0 & 0.5  & 0.0 & 3.0         \\
    Muon ID resolution                     &  0.0 & 3.7  & 0.0 & 2.0         \\
    Muon MS resolution                     &  0.0 & 1.7  & 0.0 & 4.0         \\
\hline
    \multicolumn{5}{c}{\em Signal modelling on $\mu_\mathrm{CB}$ $\mathrm{[\%]}$} \\
\hline
    Electron and photon energy scale       &  0.1 & 0.2  & 0.2 & 0.6   \\
    Muon momentum scale                    &  0.0 & 0.03 & 0.0 & 0.03   \\
    Higgs mass                             &  \alignUnderEnDash{0.2}       &  &  &         \\
    \hline
    \multicolumn{5}{c}{\em Background modelling $\mathrm{[Events]}$}   \\
    \hline
     Spurious signal                      &   1.7 & 25     &  0.005 & 6.1 \\
     \hline \hline
  \end{tabular}
\end{table}

\subsection{Uncertainties from signal and background modelling}
\label{sec:uncshape}

The uncertainties in the lepton 
and photon momentum and energy scale and resolution
impact the modelling of the signal. 
Their impact is assessed by comparing the nominal $m_{Z\gamma}$ shape parameters with the 
$m_{Z\gamma}$ shape parameters after varying
the lepton and photon momentum and energy scale and resolution by their uncertainties.
Uncertainties in both  the  position ($\mu_\mathrm{CB}$) and width ($\sigma_\mathrm{CB}$) 
of the signal $m_{Z\gamma}$ distribution are considered.

The systematic uncertainties in the muon momentum scale and resolution were determined from
$Z\to\mu\mu$ and $J/\psi\to\mu\mu$ events using the techniques described in Ref.~\cite{Aad:2016jkr}.
At $m_H=125~\GeV$, the uncertainty in the muon momentum scale (resolution) varies 
$\sigma_\mathrm{CB}$ by up to 0.5\% (4.0\%).
In the high-mass search, the effect of the muon momentum scale (resolution) uncertainty is to change
$\sigma_\mathrm{CB}$ by up to 3.0\% (4.0\%). The typical effect of the muon momentum 
scale uncertainty is to change 
$\mu_\mathrm{CB}$ by $< 0.1\%$ of its nominal value.

The systematic uncertainties in the
electron and photon energy scale and resolution follow those in Refs.~\cite{Aad:2014nim, ATLAS-egammacalib}. The
overall energy scale factors and their uncertainties were determined using $Z\to ee$ events.
Compared to Ref.~\cite{ATLAS-egammacalib}, several systematic uncertainties were re-evaluated with 
the 13~\TeV\ data, including uncertainties related to the
observed LAr cell non-linearity, the material simulation, the intercalibration of the first and
second layer of the calorimeter, and the pedestal corrections.
At $m_H=125~\GeV$, the uncertainty in the electron and photon energy scale (resolution) results in
variation in $\sigma_\mathrm{CB}$ between 0.6\% and 3.5\% 
(1.1\% and 4.0\%) depending on the 
category. For a high-mass resonance, $\sigma_\mathrm{CB}$ varies between 
1.0\% and 4.0\% 
(4.0\% and 30\%) 
due to uncertainties
in the electron/photon momentum scale (momentum resolution).
The variation in $\mu_\mathrm{CB}$ is less than 0.2\% (0.6\%) at $m_H=125~\GeV$ (at high masses).

For the $H\to Z\gamma$ search, an additional uncertainty in the assumed Higgs mass $\mH = 125.09~\GeV$ is
added in the fit, reflecting the $0.24~\GeV$~\cite{higgs-mass} uncertainty in the measured Higgs boson mass.

The uncertainty due to the choice of background function is taken to be the signal yield 
(spurious signal) obtained
when fitting the $m_{Z\gamma}$ spectra reconstructed from background-only distributions
as discussed in Section~\ref{sec:signalBkgModel}.

\subsection{Experimental uncertainties affecting the signal efficiency and acceptance}
\label{sec:uncyield}

Experimental uncertainties affecting the signal efficiency and acceptance 
can be either correlated between all event
categories (yield uncertainties) or anticorrelated between some of the categories 
(migration uncertainties) when they are related to how the signal populates the event categories.

The uncertainty in the combined 2015+2016 integrated luminosity is 3.2\%, 
correlated between all categories. It is derived, 
following a methodology similar to that detailed in Ref.~\cite{Aaboud:2016hhf}, from a preliminary 
calibration 
of the luminosity scale using $x$--$y$ beam-separation scans performed in August 2015 and May 2016.

A variation in the pile-up reweighting of the simulation is included to cover the uncertainty 
in the ratio of the predicted and measured inelastic cross sections in the fiducial volume 
defined by $m > 13~\GeV$ where $m$ is the mass of the hadronic 
system~\cite{Aaboud:2016mmw}. The uncertainty in the signal efficiency is no more than 
0.03\% (0.2\%)
for the $H\to Z\gamma$ (high-mass resonance) search.

The uncertainties in the reconstruction, identification, isolation, and trigger efficiency measurements
for muons, electrons and photons (see Section~\ref{sec:eventSelection}) are treated as fully 
correlated between all categories. They are determined from control samples of $J/\psi\to\mu\mu$
and $Z\to\mu\mu$ for muons, $J/\psi\to ee$ and $Z\to ee$ for electrons, and $Z\to\ell\ell\gamma$,
$Z\to ee$, and inclusive photons for photons, using methods described in 
Refs.~\cite{Aad:2016jkr, ATLAS-electrons, ATLAS-photonid}.

For the $H\to Z\gamma$ search, the uncertainties in the signal efficiency from the 
photon identification and isolation are found to be no more than 1.7\% and 0.4\%, respectively. 
The uncertainties in the signal efficiency
from the electron reconstruction, identification, isolation, and trigger are
found to be no more than 0.4\%, 1.6\%, 0.2\%, and 0.1\%, respectively. 
The uncertainties in the signal efficiency from the muon
selection and trigger are determined to be no more than 1.6\% and 3.5\%, respectively.
For the high-mass search, the uncertainties in the signal efficiency from 
photon identification and isolation are found to be no more than 2.6\% and 0.6\%, respectively. 
The uncertainties in the signal efficiency
from the electron reconstruction, identification, isolation, and trigger are
found to be no more than 1.0\%, 2.6\%, 3.5\%, and 0.2\%, respectively. 
The uncertainties in the efficiency from the muon
selection and trigger are determined to be as large as 0.7\% and 4.2\%, respectively.
The uncertainty due to the limited size of the simulated event samples ranges from 1.2\% to 2.0\% 
for the search for high-mass resonances.

In the $H\to Z\gamma$ search, the expected signal yield in the VBF category is 
affected by the jet energy scale and resolution and the jet vertex tagging efficiency. The corresponding 
uncertainties are anticorrelated with the other categories. Uncertainties
in the jet energy scale and resolution are estimated from 
the transverse momentum balance in dijet, $\gamma+$jet and $Z$+jet 
events~\cite{Aaboud:2017jcu}.
Uncertainties in the efficiency of the jet vertex tagging are estimated by shifting the associated 
corrections
applied to the simulation by an amount allowed by the data. The uncertainties in the category 
acceptances are as large as 4.6\%, 6.9\%, and 4.8\% from the data-driven jet calibration, the 
impact
of the jet flavour composition on the calibration, and the jet vertex tagging.

\subsection{Theoretical and modelling uncertainties}
\label{sec:unctheo}

\begin{table}
  \caption{The main sources of theoretical and modelling uncertainties for the $H\to Z \gamma$ 
    search. 
    For the uncertainties in the total efficiency and the acceptance of the different categories, 
    the gluon--gluon
    fusion samples produced with \textsc{Powheg Box} v1 with and without MPI are used, as well as 
    the nominal 
    \textsc{Powheg Box} v2 gluon--gluon fusion signal sample along with the sample
    generated with \textsc{MadGraph5\_aMC@NLO}, as described in the text.
    The combined uncertainty on the total cross section and efficiency is given assuming the
    cross sections predicted by the SM.
    The ranges for the uncertainties cover the variations among different categories.
    The uncertainty values are given as relative uncertainties.}
\label{tab:thsyst}
  \centering
  \begin{tabular}{lr@{--}l@{}}
    \hline
    \hline       
\multicolumn{1}{c}{Sources} & \alignUnderEnDash{ } & \\
\hline
     \hline

     \multicolumn{3}{c}{\em Total cross section and efficiency $\mathrm{[\%]}$} \\
     \hline
     Underlying event                         & \alignUnderEnDash{5.3} & \\
     ggF perturbative order                   & \alignUnderEnDash{3.9} & \\
     ggF PDF and $\alpha_\mathrm{s}$          & \alignUnderEnDash{3.2} & \\
     VBF perturbative order                   & \alignUnderEnDash{0.4} & \\
     VBF PDF and $\alpha_\mathrm{s}$          & \alignUnderEnDash{2.1} & \\
     $WH$ ($ZH$) perturbative order           & \alignUnderEnDash{0.5~(3.8)} & \\
     $WH$ ($ZH$) PDF and $\alpha_\mathrm{s}$  & \alignUnderEnDash{1.9~(1.6)} & \\
     Interference                             & \alignUnderEnDash{5.0} & \\
     $B$($H\to Z\gamma$)            & \alignUnderEnDash{5.9} & \\\hline
     Total (total cross section and efficiency) & \alignUnderEnDash{10} & \\
     \hline
    \multicolumn{3}{c}{\em Category acceptance $\mathrm{[\%]}$}  \\
     \hline
     ggF $H+2$-jets in VBF-enriched category  & 0.5 & 45 \\
     ggF BDT variables                        & 0.2 & 15 \\
     ggF Higgs \pt                            & 8.4 & 22 \\
     PDF and $\alpha_\mathrm{s}$              & 0.2 & 2.0 \\
     Underlying event                         & 2.9 & 25 \\\hline
     Total (category acceptance)              & 9.5 & 49\\
     \hline \hline
  \end{tabular}
\end{table}

For the $H\to Z\gamma$ search, theoretical and modelling uncertainties in the 
SM predictions for
Higgs boson production and the decay to the $Z\gamma$ final state are taken into account and
are summarised in Table~\ref{tab:thsyst}. 
They fall into
two classes: uncertainties in the total predicted cross sections, the predicted decay 
branching ratio and the total efficiency, correlated between all categories; 
and uncertainties in the event fractions per category, anticorrelated between certain
categories.

Uncertainties related to the total acceptance and efficiency for $H\to Z\gamma$ events
affect the extraction of the signal strength, the branching ratio of $H\to Z\gamma$
assuming SM Higgs boson production, as well as the product of the Higgs boson production
cross section and the branching ratio of $H\to Z\gamma$ (see Section~\ref{sec:statistics}).
The uncertainty in the total efficiency due to the modelling of multiple-parton 
interactions
is estimated from the difference in efficiency with and without multiple-particle interactions 
for the gluon--gluon fusion simulation sample, and found to be 5.3\%.

The uncertainties related to the predicted Higgs boson production cross section affect
the extraction of the signal strength as well as the branching ratio of $H\to Z\gamma$
assuming SM Higgs boson production.
The uncertainties in the predicted total cross sections of the different Higgs 
boson production processes due to the perturbative order of the calculation
and the combined uncertainties in the PDFs and $\alpha_\mathrm{s}$ are 3.9\% and 3.2\% for gluon--gluon fusion
production, respectively, 
and range from 0.4\% to 3.8\% for the other production processes for a Higgs boson mass of 125.09~\GeV\ and
a centre-of-mass energy of 13~\TeV~\cite{deFlorian:2016spz}.  
An additional $5.0\%$~\cite{Dicus:2013lta} uncertainty accounts for the effect, in the
selected phase space of the $\ell\ell\gamma$ final state, of the
interfering $H\to\ell\ell\gamma$ decay amplitudes
that are neglected in the
calculation of Ref.~\cite{deFlorian:2016spz}.
They originate from internal
photon conversion in Higgs boson decays to diphotons
($H\to\gamma^*\gamma\to\ell\ell\gamma$) or from
Higgs boson decays to dileptons with an off-shell lepton
($H\to\ell\ell^*\to\ell\ell\gamma$)~\cite{Chen:2012ju,Firan:2007tp}. 

The uncertainty in the predicted Higgs boson branching ratio to $Z\gamma$ affects the
extraction of the signal strength. 
The relative theoretical uncertainty in the predicted
Higgs boson branching ratio is 5.9\%~\cite{deFlorian:2016spz}.

Uncertainties in the modelling of kinematic distributions in the simulation of Higgs boson 
production processes affect the predicted event fractions in the different categories.
The uncertainty in the modelling of the production of jets in gluon--gluon fusion production due to
the perturbative order in QCD is estimated by scale variations in MCFM~\cite{Campbell:2010ff}. It
accounts for the uncertainty in the overall normalisation of $H+2$-jets events as well as the 
uncertainty due to the use of $\Delta\phi_{Z\gamma,jj}$, which serves to apply an implicit third-jet
veto, in the VBF BDT. The estimation of this uncertainty uses 
an extension of the Stewart--Tackmann method~\cite{Stewart:2011cf, Gangal:2013nxa}. 
It corresponds to 45\%
of the ggF contribution to the VBF category.
Additional uncertainties are assigned to account for potential mismodelling of the variables that serve as input
to the VBF BDT (see Section~\ref{sec:eventSelection}). They are estimated by reweighting the simulated ggF
events to match the distributions in $m_{jj}$, $\Delta\eta_{jj}$, $\ptt$, $\Delta R^\mathrm{min}_{Z/\gamma, j}$, 
and $|\eta_{Z\gamma}-(\eta_{j1} + \eta_{j2})/2|$ obtained from \textsc{MadGraph5\_aMC@NLO}. The resulting uncertainty
in the ggF contribution to the VBF category is 15\%.
Uncertainties in the modelling of the Higgs boson \pt\ spectrum are taken to be the envelope of the
variations of the renormalisation, factorisation, and resummation scales 
obtained using \textsc{HRes} 2.3~\cite{Grazzini:2013mca} to simulate the \pt\ spectrum.
The resulting uncertainties are evaluated using 
the Stewart--Tackmann method~\cite{Stewart:2011cf, Gangal:2013nxa} for the high relative $\pt$ and the \ptt\ categorisation
and found to range from 8.4\% to 22\%. 
Uncertainties from the choice of PDF set and $\alpha_\mathrm{s}$ are evaluated using the combined error PDF
set, which takes into account 30 variations of NNLO PDFs and two variations of $\alpha_\mathrm{s}$, 
following the PDF4LHC recommendations~\cite{Butterworth:2015oua} and are found to range from 
0.2\% to 2.0\%.
The uncertainty in the acceptance due to the modelling of multiple-parton interactions is estimated 
from the difference in acceptance with and without multiple-particle interactions for the gluon--gluon
fusion simulation sample and ranges from 2.9\% to 25\%.

\section{Statistical procedure}
\label{sec:statistics}
A profile-likelihood-ratio test statistic~\cite{stat} is used to search for
a localised excess over a smoothly falling background in the $m_{Z\gamma}$ distribution
of the data, as well as to quantify its significance and estimate either its
production cross section or signal strength.

The extended unbinned likelihood function $\mathcal{L}(\alpha, \boldsymbol{\theta})$ is given by the
product of a Poisson term, constructed from the number of observed events, $n$,
the expected event yield, $N$, and the probability density function
 of the invariant mass distribution for each candidate event $i$, 
$f_\mathrm{tot}(m_{Z\gamma}^{i}, \alpha,\boldsymbol{\theta})$~\cite{Aaboud:2016trl}:

\beq
\LL\left((\alpha, \boldsymbol{\theta})\Big|\{m_{Z\gamma}^i\}_{i=1..n}\right) = \frac{\mathrm{e}^{-N(\alpha, \boldsymbol{\theta})}N^n(\alpha, \boldsymbol{\theta})}{n!} \prod\limits_{i=1}^nf_\mathrm{tot}(m_{Z\gamma}^i,\alpha,\boldsymbol{\theta}) \times G(\boldsymbol{\theta}),
\eeq
where $\alpha$ represents the parameter of interest and 
$\boldsymbol{\theta}$ are the nuisance parameters. The function $G(\boldsymbol{\theta})$ 
represents the prior constraints on the nuisance parameters.
The expected event yield $N$ is the sum 
of the expected number of signal ($N_{\mathrm{sig}}$), background ($N_{\mathrm{bkg}}$), and
spurious signal ($\boldsymbol{N_\mathrm{spur}}\cdot\boldsymbol{\theta_\mathrm{spur}}$) events, where
$\boldsymbol{\theta_\mathrm{spur}}$ is the nuisance parameter associated with the spurious signal.
For the high-mass resonance search, the parameter of interest is $\alpha=\sigxbr$. 
The $H\to Z\gamma$ search is performed to extract several parameters of interest:
the signal strength $\mu = \sighbr / (\sighbrsm)$,
\sighbr, and \brhzg~assuming $\sigma(pp\to H)_\mathrm{SM}$. 
The signal strength $\mu$ is related to the number of signal events by
$N_{\mathrm{sig}} = L_\mathrm{tot} \times \mu \times (\sighbrsm) \times \varepsilon$, where 
$L_\mathrm{tot}$ is the total integrated luminosity, $\varepsilon$
is the signal efficiency, and  \sighbrsm~is predicted by the SM.
The theoretical uncertainties are
taken into account as described in Section~\ref{sec:sys}.

The probability density function for the invariant mass
($f_\mathrm{tot}(m_{Z\gamma}^{i}, \alpha,\boldsymbol{\theta})$) is built from the probablity 
density functions $f_\mathrm{sig}$ and $f_\mathrm{bkg}$
describing the signal and background invariant mass distributions, respectively:

\begin{multline}
f_\mathrm{tot}(m_{Z\gamma}^i,\alpha,\boldsymbol{\theta}) =
\frac{1}{N} \sum_{c} \left\{ \left[ N_{\mathrm{sig}}^{(c)}(\alpha, \boldsymbol{\theta_\mathrm{sig}}) + N_{\mathrm{spur}}^{(c)} \cdot \theta_{\mathrm{spur}}^{(c)} \right] \times f_{\mathrm{sig}}^{(c)}(m_{Z\gamma}^i,\boldsymbol{\theta_\mathrm{sig}}) \right. \\
\left.  + N_{\mathrm{bkg}}^{(c)} \times f_{\mathrm{bkg}}^{(c)}(m_{Z\gamma}^i,\boldsymbol{\theta_\mathrm{bkg}})\right\}.
\end{multline}
The index $c$ indicates the category. The $\boldsymbol{\theta_\mathrm{bkg}}$ are nuisance
parameters that determine the shape of the background.
The nuisance parameters associated with the uncertainties
in the signal parameterisation, efficiency and acceptance are denoted by 
$\boldsymbol{\theta_\mathrm{sig}}$. 
Nuisance parameters associated with uncertainties in the event yield or the
 $m_{Z\gamma}$ resolution are assigned log-normal probability density functions, while nuisance 
parameters
associated with the $m_{Z\gamma}$ signal peak position are assigned Gaussian probability 
density functions.  
The nuisance parameters associated 
with the spurious signal, $\boldsymbol{\theta_\mathrm{spur}}$, are assigned 
Gaussian probability density functions. 

The probability that the background can produce a fluctuation greater than or equal to an
excess observed in data 
is quantified by the $p$-value of the $\alpha = 0$ hypothesis,
$p_{0}$, which can also be expressed in terms of number of Gaussian standard deviations and 
provides an estimate of the local significance of a possible deviation from the expected
background. The global significance, corrected for the effect that a deviation can occur
anywhere in the search region, is estimated taking into account the trial 
factors~\cite{Gross:2010qma}.
The compatibility between the data and increasing
non-zero values of $\alpha$ is used to set upper limits at the 95\% CL on
\sighxbr, \brhzg, and the signal strength for $pp\to H\to Z\gamma$, respectively, 
using a modified frequentist 
(CL$_\mathrm{s}$) method~\cite{cls},
by identifying the value $\alpha$ for which the value of CL$_\mathrm{s}$ is equal to 0.05.

The results are derived using closed-form asymptotic formulae~\cite{stat} for masses up to
1.6~\TeV. Due to
the small number of events at large $m_{Z\gamma}$ in the high-mass resonance search, 
the results for $m_X> 1.6$~\TeV\ are derived using ensemble tests. 
The expected cross-section limits obtained from the
asymptotic formulae agree to better than 10\% with those obtained from the ensemble tests
up to $m_X$ of 1.7~\TeV.
At high $m_X$, the observed (expected) central values are underestimated by 35\% (23\%) in the 
asymptotic approach.

\section{Results}
\label{sec:result}
No evidence of a localised excess is visible near the anticipated
Higgs mass $\mH=125.09~\GeV$, 
as shown in Figure~\ref{fig:mllg-low-mass} where the invariant
mass distributions $m_{Z\gamma}$ for the individual categories of the $H\to Z\gamma$ search are
displayed with the background-only fit performed in the range of $115~\GeV < m_{Z\gamma} < 150$~\GeV.
At $\mH = 125.09$~\GeV, the observed $p$-value is \lowmassmHpz\ under the background-only hypothesis, 
in which the dominant contribution
comes from the $\mu\mu$ low \ptt category. The $p$-value corresponds to a local significance of 
\lowmassmHsignificance. 
The expected $p$-value for a SM Higgs boson at $\mH=125.09~\GeV$ is \lowmassmHpzExpSMHiggs, corresponding
to a significance of \lowmassmHsignificanceExpSMHiggs.
The observed 95\% CL limit on \sighbr~ 
is found to be \mhObserved~times the SM prediction, corresponding to
the limit of \mhObservedxsBr. Assuming SM Higgs boson production, the upper limit on
the branching ratio of the Higgs boson decay to $Z\gamma$ is found to be \mhObservedBr.
The expected 95\% CL limit on \sighbr~ assuming no (a SM) Higgs boson decay to $Z\gamma$ is 
\mhExpected\ (\mhExpectedSMHiggs) times the SM prediction.

\begin{figure}
\begin{center}
\subfigure[]{\includegraphics[width=0.375\textwidth]{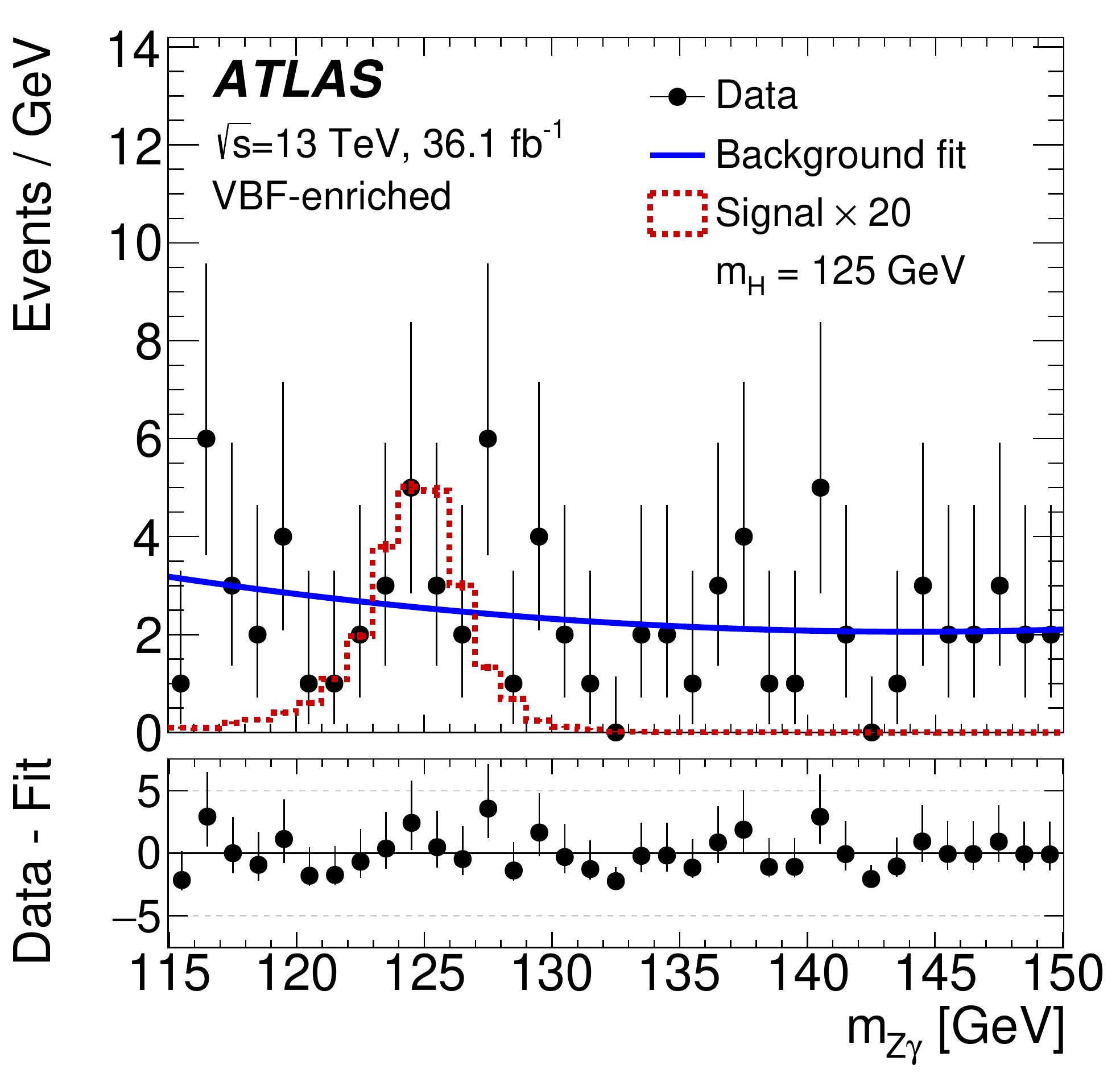}}%
\subfigure[]{\includegraphics[width=0.375\textwidth]{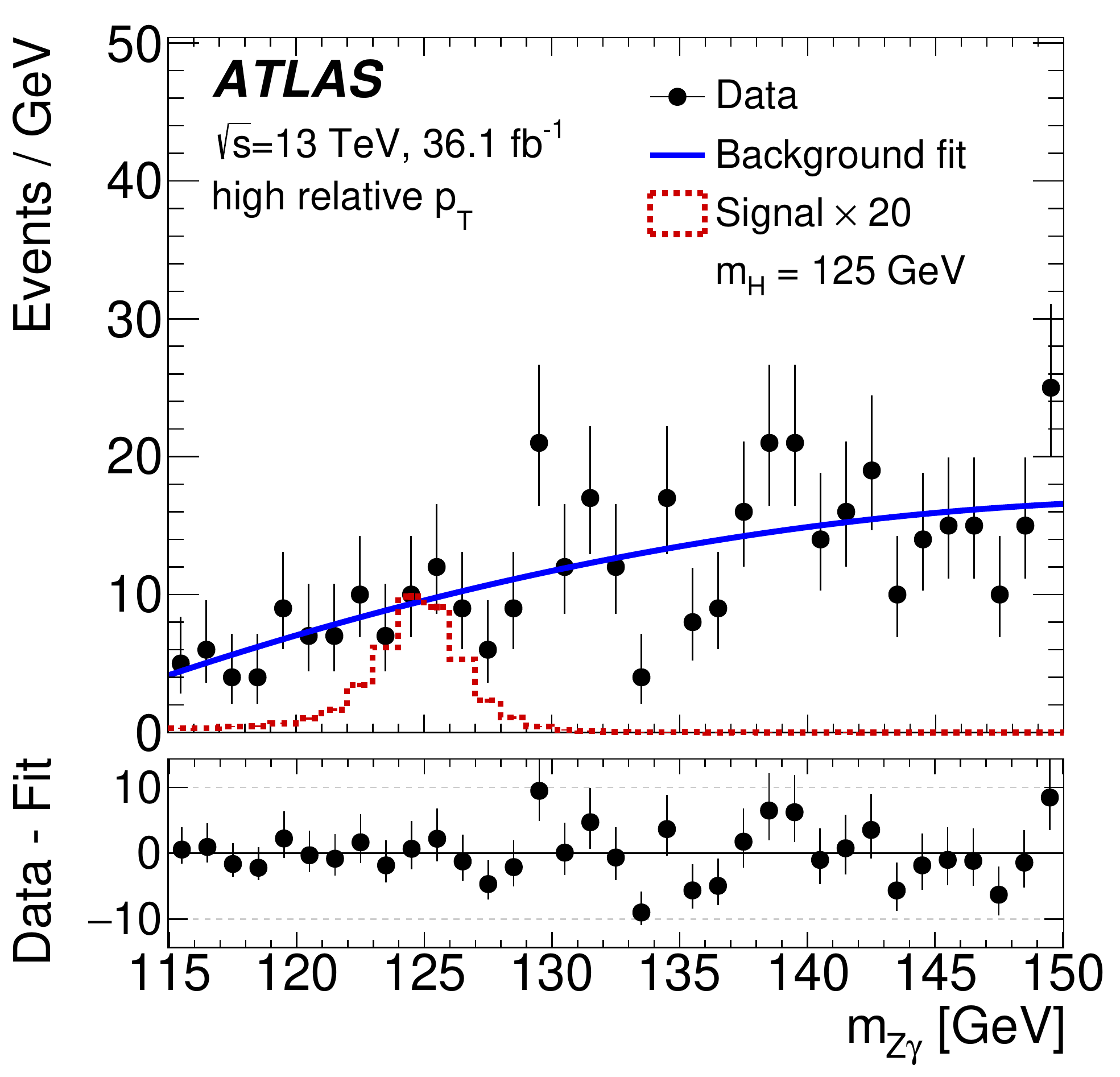}}
\subfigure[]{\includegraphics[width=0.375\textwidth]{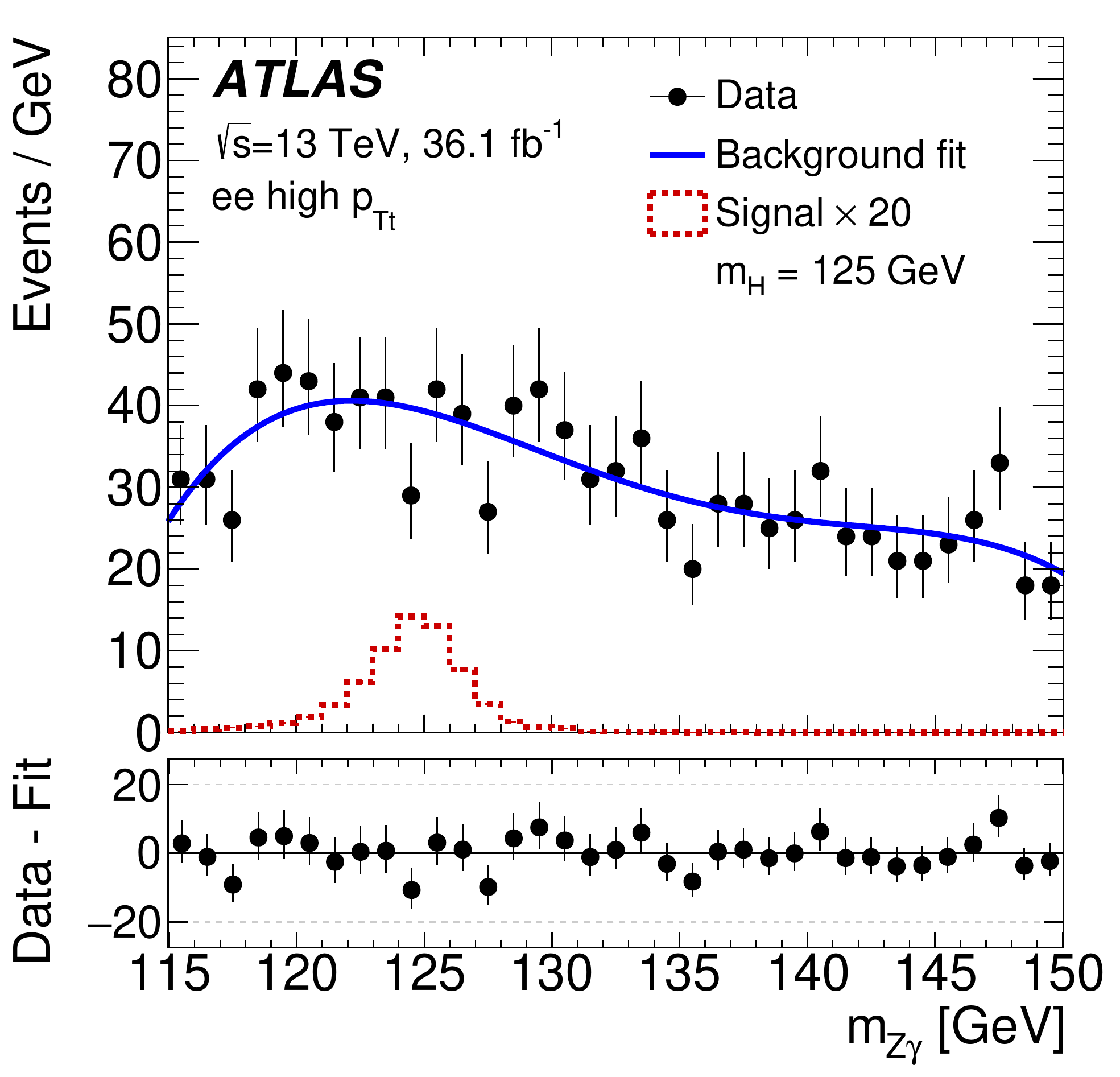}}%
\subfigure[]{\includegraphics[width=0.375\textwidth]{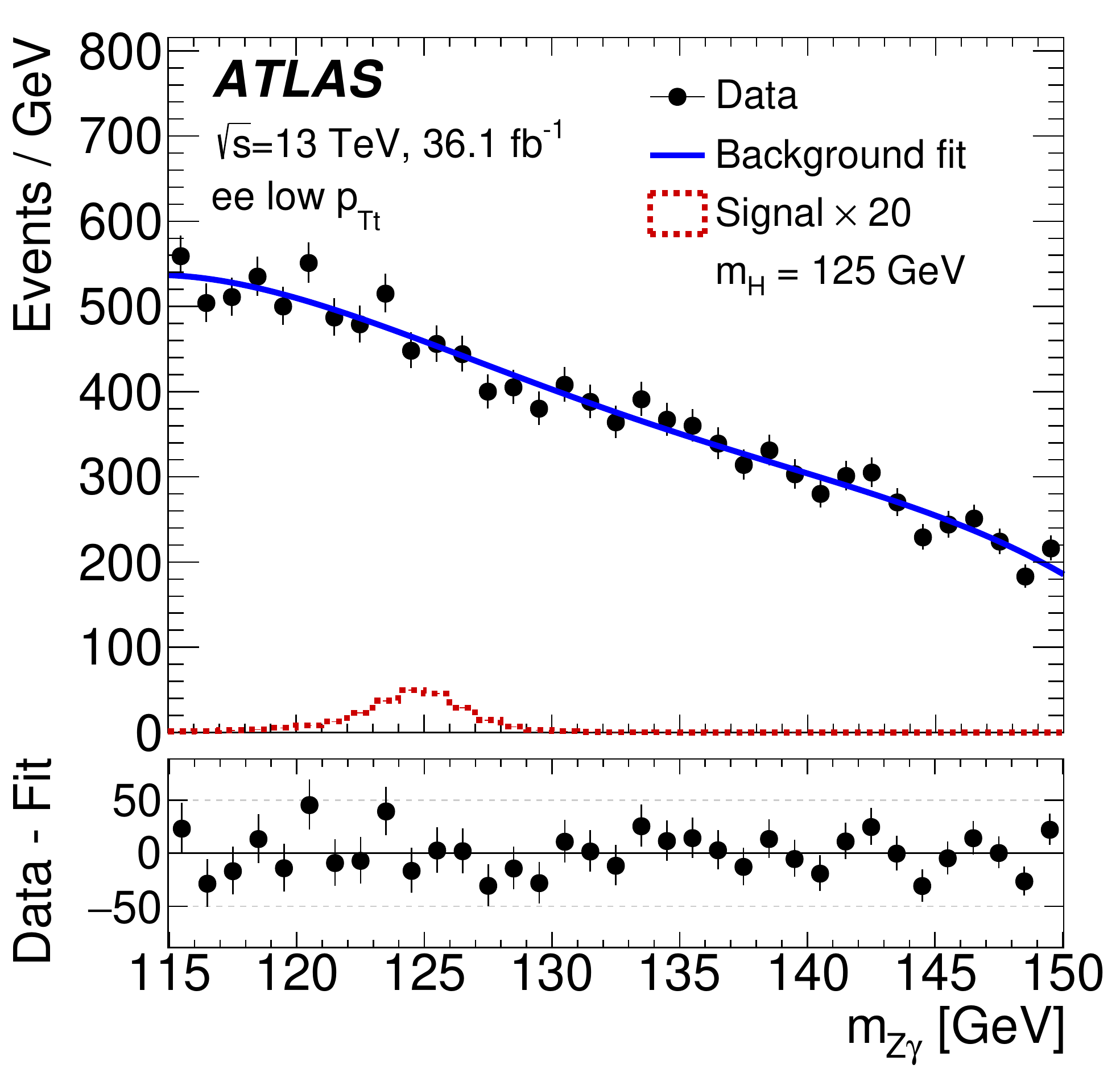}}
\subfigure[]{\includegraphics[width=0.375\textwidth]{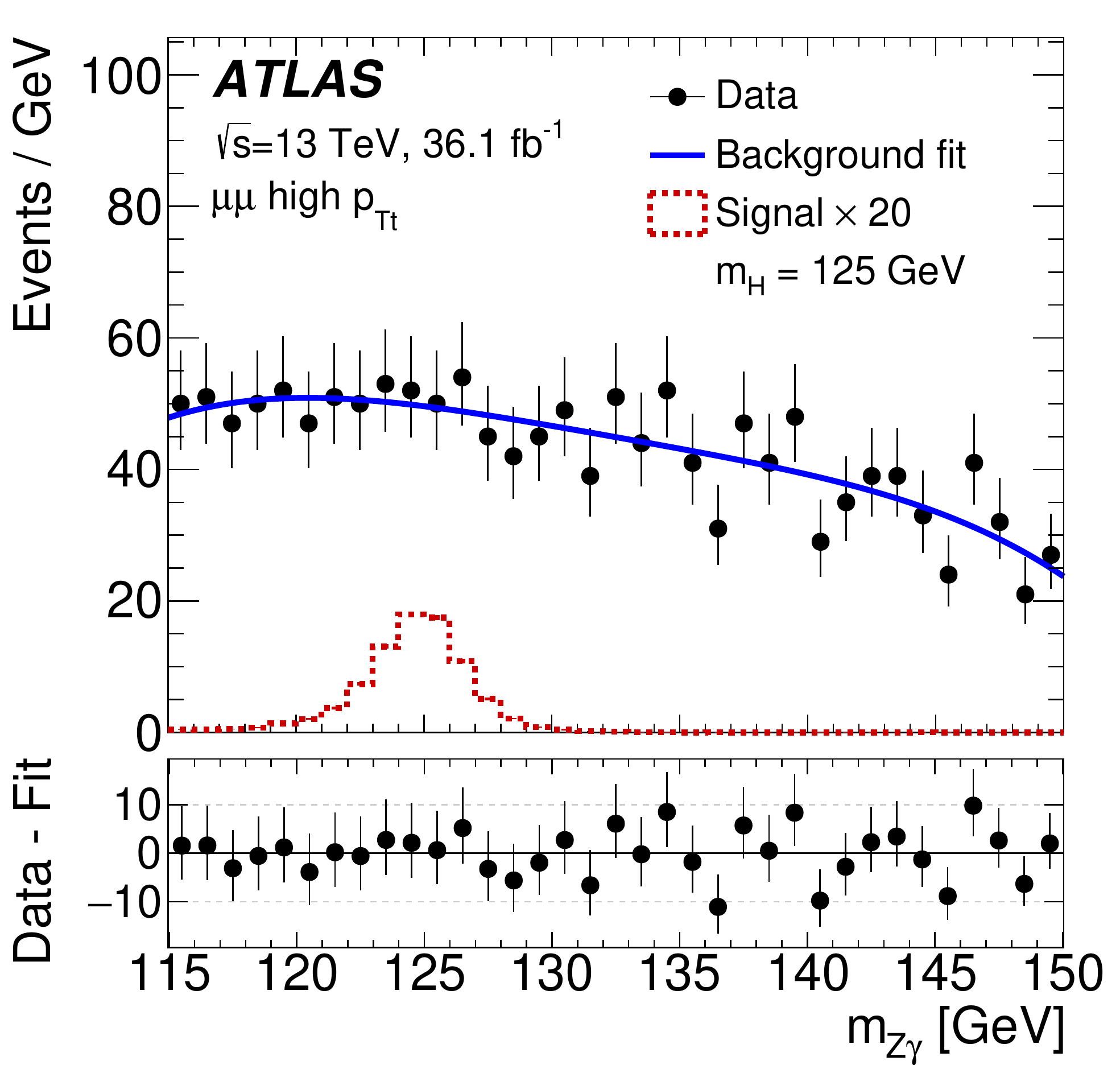}}%
\subfigure[]{\includegraphics[width=0.375\textwidth]{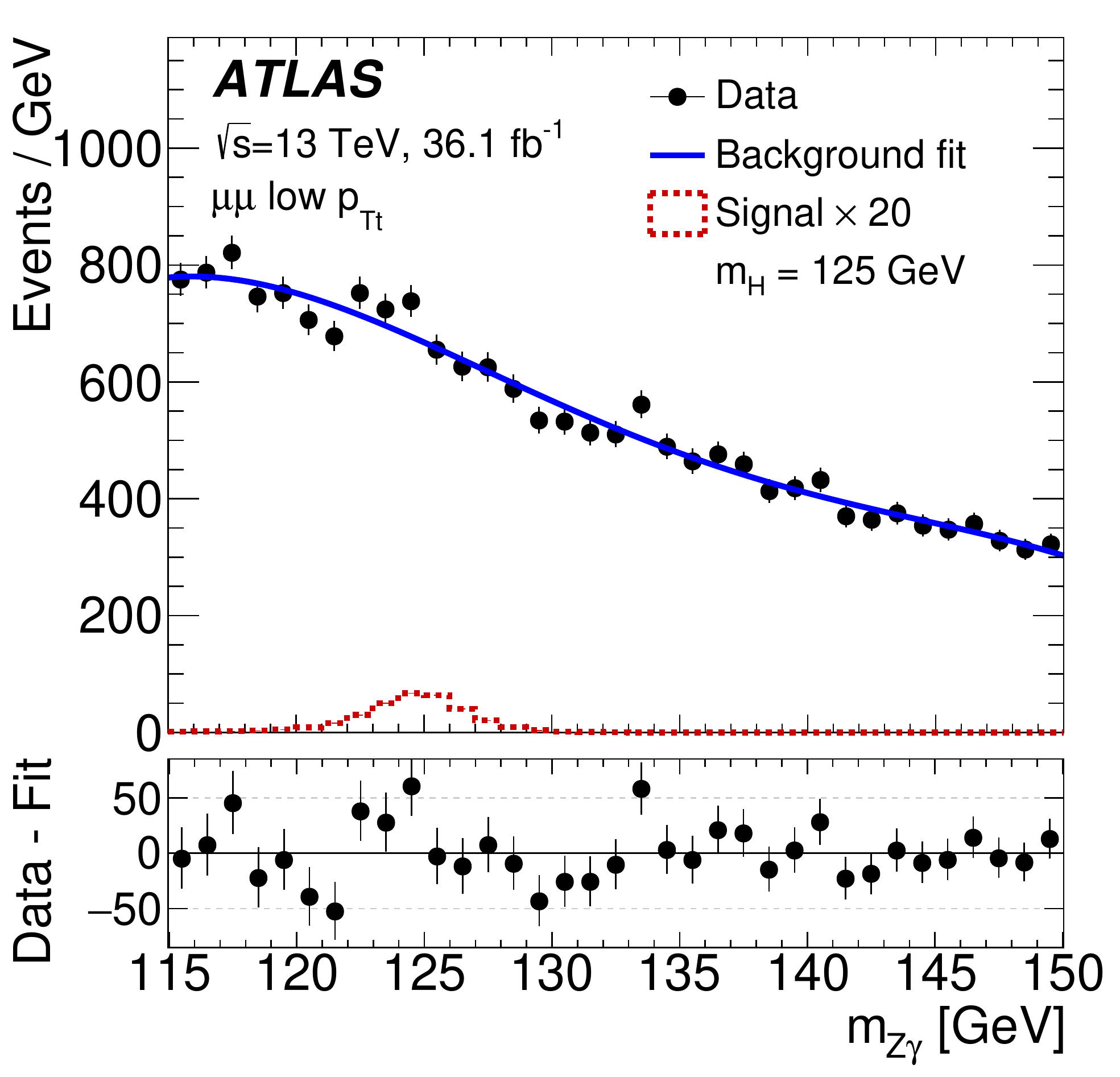}}
\end{center}
\caption{The invariant $Z\gamma$ mass ($m_{Z\gamma}$) distributions of events satisfying the 
$H\to Z\gamma$ selection in data for the six event categories:
(a) VBF-enriched, (b) high \ptg, (c) $ee$ high \ptt,  
(d) $ee$ low \ptt, (e) $\mu\mu$ high \ptt, and (f) $\mu\mu$ low \ptt.
The points represent the data and the statistical uncertainty. The solid lines show the 
background-only fits to the data, performed independently in each category.
The dashed histogram corresponds to the expected signal
for a SM Higgs boson with $\mH = 125$~\GeV\ 
decaying to $Z\gamma$ with a rate 20 times the SM prediction.
The bottom part of the figures shows the residuals of the data with respect to the background-only fit.
}
\label{fig:mllg-low-mass}
\end{figure}

The invariant mass distributions of events satisfying the 
high-mass selection are
displayed for both categories ($ee$ and $\mu\mu$) in Figure~\ref{fig:mllg-high-mass} and compared to the
background-only fit performed in the fit range $200~\GeV < m_{Z\gamma} < 2500$ \GeV. 
The highest masses measured in the \eeg\ and \mmg\ final states are
1.47~\TeV\ and 1.57~\TeV, respectively.
In the fit range, no significant excess is 
observed with respect to the background-only hypothesis. 
The largest deviation is observed around $\mX = \highmassresonance$ \GeV\
corresponding to a local significance of $\highmasshighestsignificance$.
The global significance, evaluated using the search region of 
[250--2400] \GeV\ in mass, is found to be $\highmasshighestsignificanceglobal$.
The observed and expected upper limits on \sigxbr~as a function of \mX\ are shown in Figure~\ref{fig:limit-high-mass}.
The observed limits vary between \highmasslimitlowmX\ 
and \highmasslimithighmX~for the mass range from 250~\GeV\ to 2.4~\TeV\
at the 95\% CL for a spin-0 resonance produced via gluon--gluon fusion, 
while the expected limits range from 
\highmasslimitlowmXexp\
to \highmasslimithighmXexp\ in the same mass range. For a spin-0 resonance produced via vector-boson fusion, the
limits are up to 4\% lower due to the slightly larger efficiency for this production process.

\begin{figure}
\begin{center}
\subfigure[]{\includegraphics[width=.50\textwidth]{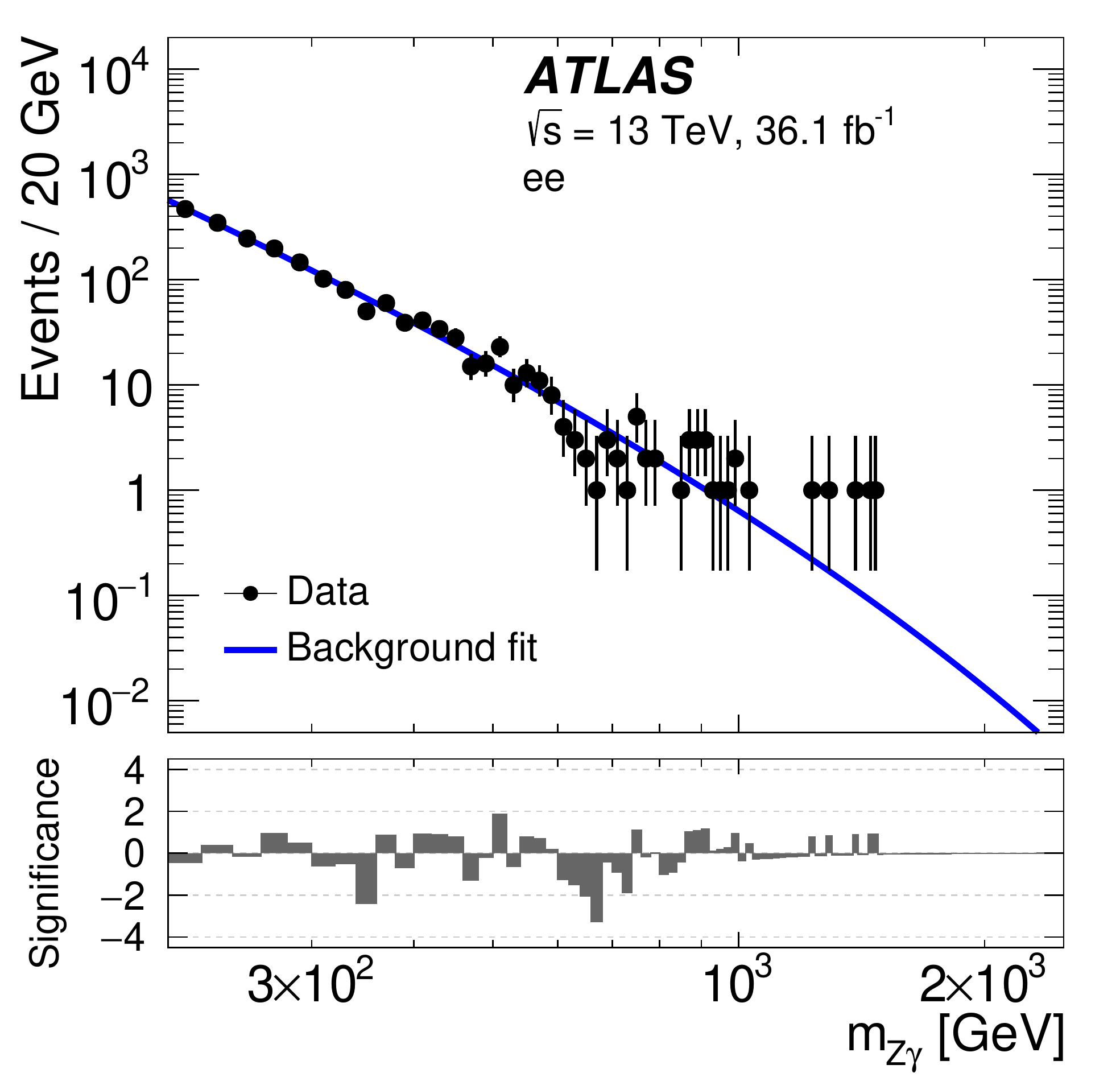}}%
\subfigure[]{\includegraphics[width=.50\textwidth]{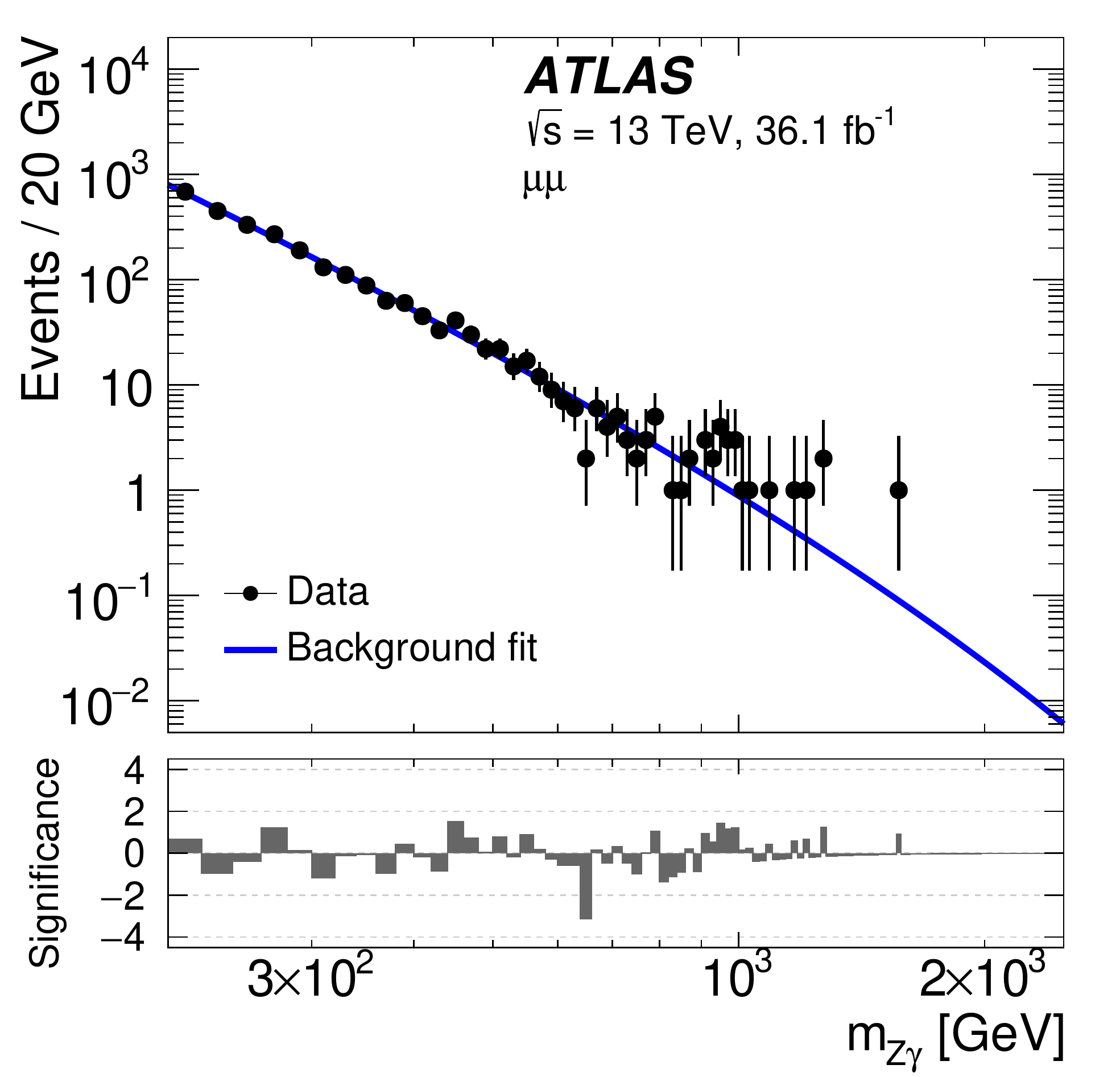}}
\end{center}
\caption{The invariant $Z\gamma$ mass ($m_{Z\gamma}$) distributions of events satisfying the 
high-mass selection in data for the two event categories:
(a) $ee$ and (b) $\mu\mu$.
The points represent the data and the statistical uncertainty. The solid lines show the
background-only fit to the data, performed independently in each category. 
The bottom part of the figures shows the significance, here defined as the residual of the data with 
respect to 
the background-only fit divided by the statistical uncertainty of the data.}
\label{fig:mllg-high-mass}
\end{figure}

\begin{figure}
\begin{center}
\includegraphics[width=.65\textwidth]{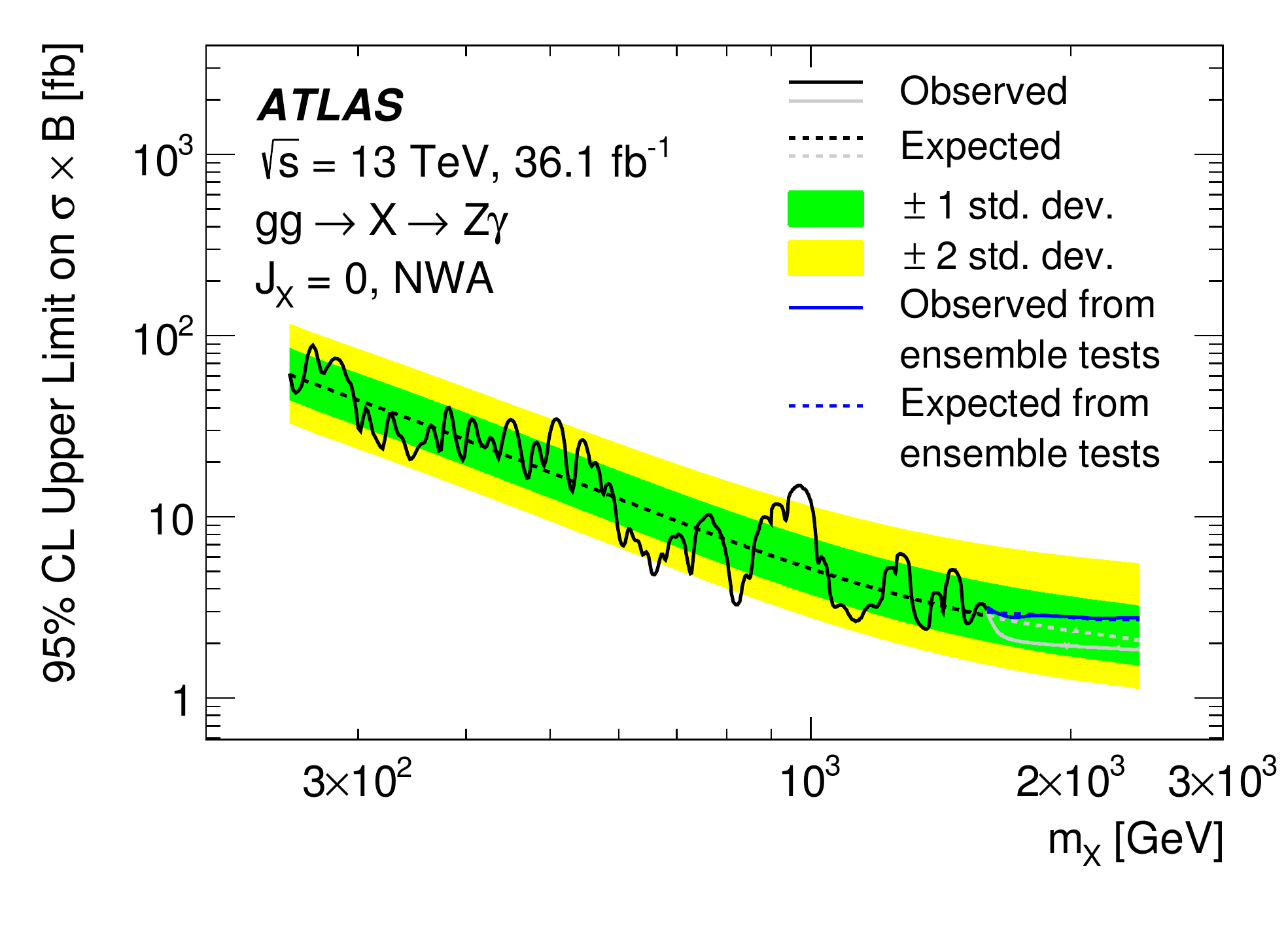}
\end{center}
\caption{The observed (solid line) and expected (dashed line) upper limit derived at the 95\% CL
on \sigxbr~at \sqtt~as a function of the high-mass spin-0 resonance's mass, 
assuming production via gluon--gluon fusion and using the narrow width assumption (NWA).
For $m_X> 1.6$~\TeV\ results are derived from ensemble tests in addition to the results obtained 
using closed-form asymptotic formulae.
The shaded regions correspond to the $\pm 1$ and $\pm 2$ standard deviation bands for the expected 
exclusion limit derived using asymptotic formulae.}
\label{fig:limit-high-mass}
\end{figure}

The results are also interpreted in terms of spin-2 resonances. Observed and expected upper 
limits at the 95\% CL
on \sigxbr~are derived and shown in Figure~\ref{fig:limit-spin2-high-mass} for both the $gg$
and $q\bar{q}$ processes. 
The observed limits for the $gg$ ($q\bar{q}$) process
vary between \highmasslowlimitgg~(\highmasslowlimitqq) 
and \highmasshighlimitgg~(\highmasshighlimitqq) for the mass range from 250~\GeV\ to 2.4~\TeV,
while the expected limits range between \highmasslowlimitggexp~(\highmasslowlimitqqexp)
and \highmasshighlimitggexp~(\highmasshighlimitqqexp) in the same mass range.

\begin{figure}
\begin{center}
\subfigure[]{\includegraphics[width=.50\textwidth]{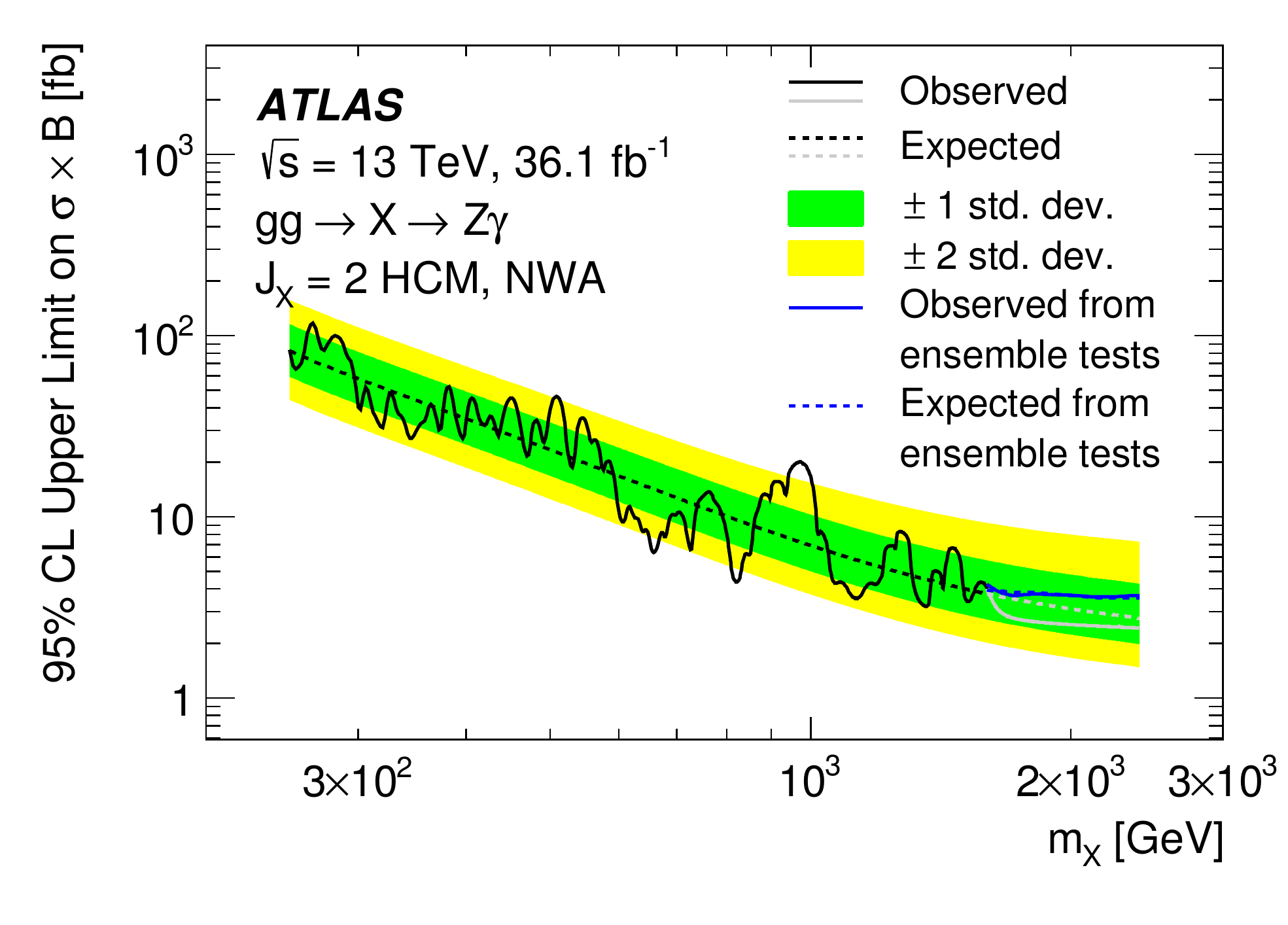}}%
\subfigure[]{\includegraphics[width=.50\textwidth]{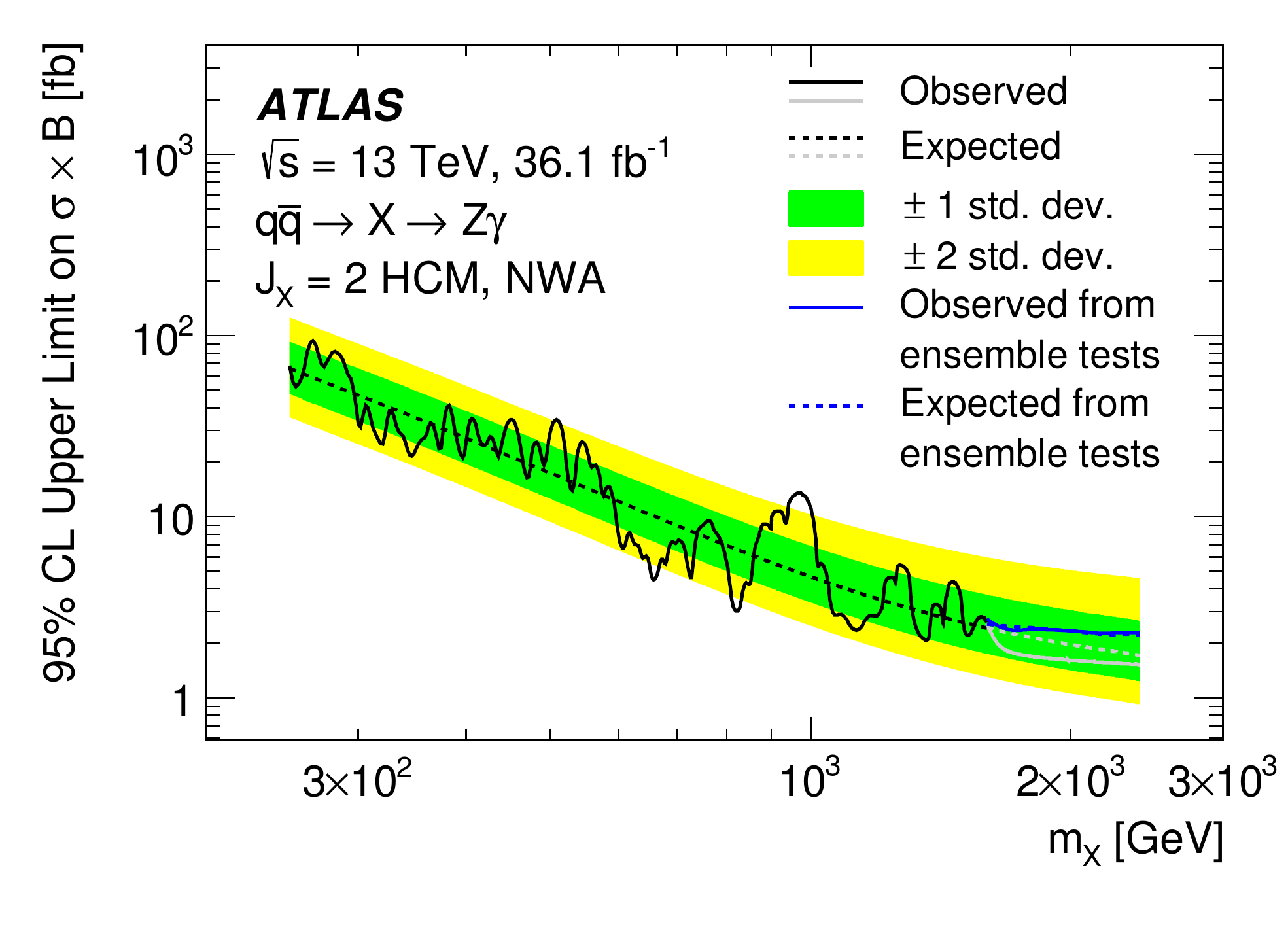}}
\end{center}
\caption{The observed (solid line) and expected (dashed line) upper limit derived at the 95\% CL
on \sigxbr~at \sqtt~as a function of the spin-2
resonance mass produced via (a) gluon--gluon initial states and 
(b) $q\bar{q}$ initial states modelled using the Higgs Characterisation Model (HCM), using
the narrow width assumption (NWA).
For $m_X> 1.6$~\TeV\ results are derived from ensemble tests in addition to the results obtained 
using closed-form asymptotic formulae.
The shaded regions correspond to the $\pm 1$ and $\pm 2$ standard deviation 
bands for the expected exclusion limit derived using asymptotic formulae.}
\label{fig:limit-spin2-high-mass}
\end{figure}

The limits on \sigxbr\ for high-mass resonances are valid for resonances with a natural width that is
small compared to the detector resolution.

\FloatBarrier

\section{Conclusion}
\label{sec:conclusion}
Searches for \Zg decays of the SM Higgs boson (\htoZg) and of a narrow high-mass resonance (\xtoZg) in 
\lumithirteentev~\invfb of $pp$ collisions at \sqtt~at 
the LHC have been performed with the ATLAS experiment.
The observed data are consistent with the
expected background. No evidence for the \htoZg~and \xtoZg~decays is observed and upper limits
 are set on \sighbr~at $\mH =125.09$~\GeV\ and also on \sigxbr~as a function of $m_X$. 
For the Higgs mass of 125.09~\GeV, the observed 95\% CL upper
limit on the \sighbr~is \mhObserved~times the SM prediction.
The search for high-mass $Z\gamma$
resonances was studied using both the spin-0 and spin-2 interpretations.
The observed limit varies between \highmasslimitlowmX\
and \highmasslimithighmX\ for the mass range from 250~\GeV\ to 2.4~\TeV\
for a spin-0 resonance, where a resonance produced in gluon--gluon fusion is used
as a benchmark model. 
For spin-2 resonances, LO predictions from the Higgs Characterisation Model are used as
benchmarks. The limits for spin-2 resonances range
between \highmasslowlimitgg~(\highmasslowlimitqq)
and \highmasshighlimitgg~(\highmasshighlimitqq) for the the mass range from 250~\GeV\ to 2.4~\TeV\ for a
resonance produced via gluon--gluon (quark--antiquark) initial states. 
The corresponding expected limits for this mass range vary between \highmasslimitlowmXexp\
and \highmasslimithighmXexp\ for a spin-0 resonance, and 
between \highmasslowlimitggexp~(\highmasslowlimitqqexp)
and \highmasshighlimitggexp~(\highmasshighlimitqqexp) for a spin-2
resonance produced via gluon--gluon (quark--antiquark) initial states.

\section*{Acknowledgements}
\label{sec:acknowledgements}

We thank CERN for the very successful operation of the LHC, as well as the
support staff from our institutions without whom ATLAS could not be
operated efficiently.

We acknowledge the support of ANPCyT, Argentina; YerPhI, Armenia; ARC, Australia; BMWFW and FWF, Austria; ANAS, Azerbaijan; SSTC, Belarus; CNPq and FAPESP, Brazil; NSERC, NRC and CFI, Canada; CERN; CONICYT, Chile; CAS, MOST and NSFC, China; COLCIENCIAS, Colombia; MSMT CR, MPO CR and VSC CR, Czech Republic; DNRF and DNSRC, Denmark; IN2P3-CNRS, CEA-DSM/IRFU, France; SRNSF, Georgia; BMBF, HGF, and MPG, Germany; GSRT, Greece; RGC, Hong Kong SAR, China; ISF, I-CORE and Benoziyo Center, Israel; INFN, Italy; MEXT and JSPS, Japan; CNRST, Morocco; NWO, Netherlands; RCN, Norway; MNiSW and NCN, Poland; FCT, Portugal; MNE/IFA, Romania; MES of Russia and NRC KI, Russian Federation; JINR; MESTD, Serbia; MSSR, Slovakia; ARRS and MIZ\v{S}, Slovenia; DST/NRF, South Africa; MINECO, Spain; SRC and Wallenberg Foundation, Sweden; SERI, SNSF and Cantons of Bern and Geneva, Switzerland; MOST, Taiwan; TAEK, Turkey; STFC, United Kingdom; DOE and NSF, United States of America. In addition, individual groups and members have received support from BCKDF, the Canada Council, CANARIE, CRC, Compute Canada, FQRNT, and the Ontario Innovation Trust, Canada; EPLANET, ERC, ERDF, FP7, Horizon 2020 and Marie Sk{\l}odowska-Curie Actions, European Union; Investissements d'Avenir Labex and Idex, ANR, R{\'e}gion Auvergne and Fondation Partager le Savoir, France; DFG and AvH Foundation, Germany; Herakleitos, Thales and Aristeia programmes co-financed by EU-ESF and the Greek NSRF; BSF, GIF and Minerva, Israel; BRF, Norway; CERCA Programme Generalitat de Catalunya, Generalitat Valenciana, Spain; the Royal Society and Leverhulme Trust, United Kingdom.

The crucial computing support from all WLCG partners is acknowledged gratefully, in particular from CERN, the ATLAS Tier-1 facilities at TRIUMF (Canada), NDGF (Denmark, Norway, Sweden), CC-IN2P3 (France), KIT/GridKA (Germany), INFN-CNAF (Italy), NL-T1 (Netherlands), PIC (Spain), ASGC (Taiwan), RAL (UK) and BNL (USA), the Tier-2 facilities worldwide and large non-WLCG resource providers. Major contributors of computing resources are listed in Ref.~\cite{ATL-GEN-PUB-2016-002}.

\printbibliography

\newpage
\begin{flushleft}
{\Large The ATLAS Collaboration}

\bigskip

M.~Aaboud$^\textrm{\scriptsize 137d}$,
G.~Aad$^\textrm{\scriptsize 88}$,
B.~Abbott$^\textrm{\scriptsize 115}$,
O.~Abdinov$^\textrm{\scriptsize 12}$$^{,*}$,
B.~Abeloos$^\textrm{\scriptsize 119}$,
S.H.~Abidi$^\textrm{\scriptsize 161}$,
O.S.~AbouZeid$^\textrm{\scriptsize 139}$,
N.L.~Abraham$^\textrm{\scriptsize 151}$,
H.~Abramowicz$^\textrm{\scriptsize 155}$,
H.~Abreu$^\textrm{\scriptsize 154}$
% ...
\bigskip
\\
% ...
$^{12}$ Institute of Physics, Azerbaijan Academy of Sciences, Baku, Azerbaijan\\
% ...
$^{88}$ CPPM, Aix-Marseille Universit{\'e} and CNRS/IN2P3, Marseille, France\\
% ...
$^{115}$ Homer L. Dodge Department of Physics and Astronomy, University of Oklahoma, Norman OK, United States of America\\
% ...
$^{119}$ LAL, Univ. Paris-Sud, CNRS/IN2P3, Universit{\'e} Paris-Saclay, Orsay, France\\
% ...
$^{137}$ $^{(a)}$ Facult{\'e} des Sciences Ain Chock, R{\'e}seau Universitaire de Physique des Hautes Energies - Universit{\'e} Hassan II, Casablanca; $^{(b)}$ Centre National de l'Energie des Sciences Techniques Nucleaires, Rabat; $^{(c)}$ Facult{\'e} des Sciences Semlalia, Universit{\'e} Cadi Ayyad, LPHEA-Marrakech; $^{(d)}$ Facult{\'e} des Sciences, Universit{\'e} Mohamed Premier and LPTPM, Oujda; $^{(e)}$ Facult{\'e} des sciences, Universit{\'e} Mohammed V, Rabat, Morocco\\
% ...
$^{139}$ Santa Cruz Institute for Particle Physics, University of California Santa Cruz, Santa Cruz CA, United States of America\\
% ...
$^{151}$ Department of Physics and Astronomy, University of Sussex, Brighton, United Kingdom\\
% ...
$^{154}$ Department of Physics, Technion: Israel Institute of Technology, Haifa, Israel\\
$^{155}$ Raymond and Beverly Sackler School of Physics and Astronomy, Tel Aviv University, Tel Aviv, Israel\\
% ...
$^{161}$ Department of Physics, University of Toronto, Toronto ON, Canada\\
% ...
$^{a}$ Also at Department of Physics, King's College London, London, United Kingdom\\
$^{b}$ Also at Institute of Physics, Azerbaijan Academy of Sciences, Baku, Azerbaijan\\
$^{c}$ Also at Novosibirsk State University, Novosibirsk, Russia\\
$^{d}$ Also at TRIUMF, Vancouver BC, Canada\\
$^{e}$ Also at Department of Physics {\&} Astronomy, University of Louisville, Louisville, KY, United States of America\\
% ...
$^{*}$ Deceased
\end{flushleft}


\end{document}